\PassOptionsToPackage{dvipsnames}{xcolor}
\documentclass[nonacm]{acmart}

\usepackage[utf8]{inputenc}
\usepackage{ifthen}
\usepackage{hyperref}

\usepackage{mathtools}
\usepackage{soul}
\usepackage[normalem]{ulem}
\usepackage{graphicx}
\usepackage{pdfpages}
\usepackage{array}
\usepackage{epsf,picinpar}
\usepackage{varioref}
\usepackage{fdsymbol}
\usepackage{comment}
\usepackage{colortbl}
\usepackage{enumitem}
\usepackage{graphicx}
\usepackage{subcaption}
\usepackage{pifont}
\usepackage{fancyhdr}
\usepackage{multirow}
\usepackage{tabularray}
\usepackage{tablefootnote}
\usepackage{longtable}
\usepackage{tabularray}
\usepackage{tikz}
\usetikzlibrary{arrows.meta, positioning}
\usepackage{xfrac}
\usepackage{float} 
\usepackage{pgf-pie}

\usepackage{tikz}
\usepackage{subcaption}
\usepackage[title]{appendix}%
\usepackage{textcomp}%
\usepackage{manyfoot}%
\usepackage{booktabs}%
\usepackage{listings}%
\usepackage{tabularx}
\usepackage[framemethod=TikZ]{mdframed}
\usepackage{pgfplots}

\usepackage{algorithmic}
\usepackage[linesnumbered,lined,commentsnumbered]{algorithm2e}

\usepackage{AMMALanguages}

\usepackage{makecell}
\usepackage{adjustbox}
\newcommand{\secref}[1]{Sec.~\ref{#1}}
\newcommand{\figref}[1]{Fig.~\ref{#1}}
\newcommand{\tabref}[1]{Tab.~\ref{#1}}
\newcommand{\lstref}[1]{Listing~\ref{#1}}
\newcommand{\appref}[1]{Appendix~\ref{#1}}

\newcommand{\todo}[1]{
    \ifthenelse{\boolean{showannotations}}%
    {\ifthenelse{\equal{#1}{}}{\textcolor{red}{TODO}}{\textcolor{red}{TODO:~{#1}}}}%
    {}%
}

\newcommand{\assignedto}[1]{%
    \ifthenelse{\boolean{showannotations}}%
    {\textbf{\noindent\ding{46}\textcolor{white}{~\colorbox{\assignementcolor}{Assigned to:}}~\textcolor{\assignementcolor}{#1}\\}%
    }
    {}
}

\newcommand{\rem}[1]{%
    \ifthenelse{\boolean{showannotations}}%
    {\textcolor{oldtextcolor}{\st{#1}}}%
    {}%
}

\newcommand\add[1]{%
    \ifthenelse{\boolean{showannotations}}%
    {\textcolor{newtextcolor}{{#1}}}%
    {#1}%
}

\newcommand\rep[2]{%
    \ifthenelse{\boolean{showannotations}}%
    {\rem{#1}~\add{#2}}%
    {#2}%
}

\newenvironment{conclusionframe}[1]
  {\mdfsetup{
    frametitle={\colorbox{white}{\space#1\space}},
    innertopmargin=2pt,
    frametitleaboveskip=-\ht\strutbox,
    frametitlealignment=\center
    }
  \phantom{}  
  \begin{mdframed}%
  }
  {\end{mdframed}}

\newcommand{\smallsection}[1]{%
    
    \phantom{}
    
    \noindent\textbf{{#1}.}
}

\newcommand{\code}[1]{\texttt{#1}}

\newboolean{showannotations}
\setboolean{showannotations}{true} 

\newboolean{shownames}
\setboolean{shownames}{true}

\pgfplotsset{width=10cm,compat=1.9}

\definecolor{newtextcolor}{rgb}{0, 0, 1}
\definecolor{oldtextcolor}{rgb}{1, 0, 0}
\newcommand{\assignementcolor}{orange}
\definecolor{highlightcolor}{rgb}{.99, 1, .0}
\sethlcolor{highlightcolor}
\definecolor{darkred}{rgb}{0.6,0.0,0.0}

\setcopyright{acmcopyright}
\acmYear{2024}
\copyrightyear{2024}

\title{On the Utility of Domain  Modeling Assistance with Large Language Models}
\author{Meriem Ben Chaaben}
\orcid{0000-0000-0000-0000}
\affiliation{
  \institution{DIRO, Université de Montréal}
  \city{Montréal}
  \country{Canada}
}
\email{meriem.ben.chaaben@umontreal.ca}

\author{Lola Burgue{\~n}o}
\orcid{0000-0002-7779-8810}
\affiliation{
  \institution{ITIS Software, University of Malaga}
  \city{Malaga}
  \country{Spain}
}
\email{lolaburgueno@uma.es}

\author{Istvan David}
\orcid{0000-0002-4870-8433}
\affiliation{
  \institution{McMaster University}
  \city{Hamilton}
  \country{Canada}
}
\email{istvan.david@mcmaster.ca}

\author{Houari Sahraoui}
\orcid{0000-0001-6304-9926}
\affiliation{
  \institution{DIRO, Université de Montréal}
  \city{Montréal}
  \country{Canada}
}
\email{sahraouih@iro.umontreal.ca}

\renewcommand{\shortauthors}{Ben Chaaben et al.}

\begin{document}

\begin{abstract}

Model-driven engineering (MDE) simplifies software development through abstraction, yet challenges such as time constraints, incomplete domain understanding, and adherence to syntactic constraints hinder the design process. This paper presents a study to evaluate the usefulness of a novel approach utilizing large language models (LLMs) and few-shot prompt learning to assist in domain modeling.  
The aim of this approach is to overcome the need for extensive training of  AI-based completion models on scarce domain-specific datasets and to offer versatile support for various modeling activities, providing valuable recommendations to software modelers. To support this  approach, we developed MAGDA, a user-friendly tool, through which we conduct a user study and assess the real-world applicability of our approach in the context of domain modeling, offering valuable insights into its usability and effectiveness.  
\end{abstract}

\maketitle

\section{Introduction}\label{sec:intro}

Software modeling helps manage the complexity of engineering problems, enhancing the efficiency and effectiveness of software development through the power of abstraction~\cite{brambilla2017model, schmidt2006guest}. In particular, domain modeling is a crucial step in software design. It is defined as an explicit, structured, and visual representation of real-world objects within a domain and their interconnections~\cite{chursin2017}. These representations involve vocabulary, key concepts, behavior, and relationships.

Domain modeling plays a pivotal role in prompting a comprehensive understanding of the problem at hand and establishing the groundwork for developing solutions. However, it is a complex and error-prone task. Firstly, it necessitates proficiency in two distinct areas of expertise: domain knowledge and modeling formalisms, posing challenges for both domain experts and software specialists. Furthermore, a domain is an open world, making it highly contextual to delineate the boundaries of what should be included in the model. Lastly, the subsequent design phases heavily rely on the resulting domain models, meaning that the omission of certain concepts can significantly impact the developed solutions.

These challenges have led to an increased demand for new tools and techniques that can assist modelers in their tasks. While AI has made significant progress in code-related tasks such as code completion, program repair, and testing~\cite{rozière2024code}, its application in the early software development phases, specifically modeling and design, remains limited. More specifically, domain modeling assistance has primarily been addressed by searching for model fragments in repositories~\cite{lopez2020mar} or by exploiting formalized knowledge~\cite{agtrickauer2018domore}. More recently, deep-learning models have been trained or fine-tuned using modeling data (e.g., ~\cite{weyssow2022recommending}). However, these research initiatives were limited by the scarcity of domain-specific training data required for high-quality recommendation systems.

With the increase in the performance of Large Language Models (LLMs), new approaches based on prompting have been recently proposed to assist in domain modeling by suggesting domain concepts, i.e., automated model completion~\cite{chen2023automated}. These approaches have demonstrated good performance in the correctness of the suggested concepts, but they still have many shortcomings~\cite{CamaraTBV23}. Specifically, for a recommender system, the correctness of suggestions is not the only utility aspect that should be considered.

In this paper, we empirically study various aspects of the utility of LLM-based domain modeling assistance. The notion of utility is closely tied to recommender systems, and can be objective or perceived. In the context of modeling assistance, we focus on three key objective aspects: productivity, gauged by the time taken to complete the modeling task; contributivity, measured by the extent of suggestion contribution to the final model; and creativity, approximated by the diversity of models of a given domain created by different modelers using the same assistant. We also explore the perceived utility through the suggestion mode preference and the modeling experience.

The study was conducted with 30 subjects from two countries, Canada and Spain, using a state-of-the-art approach based on few-shot prompt learning, and its corresponding tool.
In particular, we aim to answer the following research questions when incorporating few-shot learning techniques for suggesting modeling elements. These research questions are grouped into the two utility categories as follows.

\paragraph*{Objective utility:}
\begin{description}
    \item[RQ1 (Productivity).] \textit{What is the impact of modeling assistance on the \textbf{time} required to complete the domain models?}
    \item[RQ2 (Contributivity).] \textit{What is the level of contribution of the suggestions to the resulting models?}

    \item[RQ3 (Creativity).] \textit{Does the assistance reduce the solution \textbf{diversity} among the modelers?}
\end{description}

\paragraph*{Perceived utility:}
\begin{description}
    \item[RQ4.] \textit{What kind of assistance do participants \textbf{prefer to choose} when completing their tasks?}
    \item[RQ5.] \textit{How does the assistance affect the \textbf{modeling experience}, as perceived by the modelers?}

\end{description}

Our findings indicate that LLM-based assistance significantly reduces the time required to complete modeling tasks, thereby boosting productivity. Additionally, we found that providing suggestions throughout the modeling process increases both the likelihood of their acceptance and their positive impact on the final models, compared to offering suggestions only at the end of the activity. Another observation is that LLM-based assistance slightly reduces the diversity of models produced by different participants within the same domain, which may limit modelers' creativity. However, this reduction is not substantial enough to draw definitive conclusions. Lastly, participant feedback highlights the utility of LLM-based assistance in model completion, with users expressing a preference for tools that are adaptable and intuitive.

The remainder of the paper is organized as follows. In section~\ref{sec:background} we present an exploration of assistance in software modeling as well as an overview of Large Language models and few-shot learning. We also reviews relevant existing work. Section~\ref{sec:approach}, reveals our proposed approach as well as an illustrative running example. Section~\ref{sec:tool}, details the tool we implemented for domain modeling assistance. The Evaluation methodology is presented in section~\ref{sec:studydesign} while the obtained results are discussed in section~\ref{sec:discussion}. Finally, section~\ref{sec:conclusion} concludes the paper.

\section{Background and Related work}\label{sec:background}

In this section, we provide an overview of the background concepts of our study and review the related work, focusing on assistance in software modeling (\secref{sec:background-assistance}) and large language models (\secref{sec:background-llm}).

\subsection{Assistance in Software Modeling}\label{sec:background-assistance}

The complexity of modern software systems often requires rigorous engineering methods.
Permanent examples include cyber-physical systems, especially those of safety-critical nature, such as automotive systems.
Model-driven engineering (MDE)~\cite{schmidt2006guest} advocates modeling software systems before they get released and use these models for an array of crucial engineering activities, including requirements formalization~\cite{dromey2003requirements}, rapid prototyping and validation~\cite{moeskopf2008introduction}, formal verification of functional~\cite{gonzalez2014formal} or extra-functional properties, and code generation~\cite{kelly2008domain}.
Within MDE, \emph{domain modeling}~\cite{iscoe1991domain,prieto1991domain} focuses on externalizing core domain concepts and relationships among them, typically relying on a subset of the Unified Modeling Language (UML) Class Diagram notation~\cite{berardi2005reasoning}. Class diagrams have shown to offer a balanced combination of descriptive power and economical notation, allowing for an efficient modeling experience~\cite{lindland1994understanding}. Unfortunately, domain modeling is still a time-consuming and error-prone task when carried out manually.

To alleviate the issues stemming from manual efforts, computer-aided automation has been of particular interest in software modeling~\cite{mussbacher2020opportunities}.
Traditional software modeling assistance relies on pattern-based~\cite{kuschke2014pattern} or similarity-based~\cite{elkamel2016uml} recommendations, and occasionally it relies on recommendations extracted from natural language specifications~\cite{ibrahim2010class}.
\citet{savary2023understanding} show that effective modeling assistance requires incorporating high-context information in the assistance mechanism, such as domain knowledge and corporate methodology, as well as intuitive notations and tools. Additional criteria for effective and efficient modeling assistance include usability, competence, added value, adaptation to the user, and understanding of the work context.
We focus on such human aspects when we gauge the perceived and objective utility of modeling assistance. 
While their advantages are clear, traditional model assistant techniques still suffer from limited availability of training data and the need for substantial computational resources.

AI-based techniques opened new frontiers in modeling assistance. AI-based techniques promise better assistance quality due to the more powerful patterns extracted from the training data. By extension, however, the volume of available training data becomes a crucial consideration.
\citet{weyssow2022recommending} propose a similarity-based, data-centered approach in which datasets of models and metamodels relevant to the target objective are transformed into textual representations to train neural networks.
A similarity-based approach is proposed in the work of  \citet{elkamel2016uml}.  It involves using clustering algorithms instead of neural networks to propose UML classes.
Other techniques leverage the fact that software models are eventually used for generating source code and take a reverse engineering stance. \citet{capuano2022learning} train a neural network with data extracted from Java projects.
Similarly, \citet{rocco2021gnn}, exploit large code bases to reverse-engineer to train a RoBERTa language model~\cite{liu2019roberta}. Their results show substantial bias in the obtained recommendations towards implementation-related concepts rather than high-level domain concepts, such as \textit{utilities} (cf. Util classes in Java) and \textit{controllers} (cf. Controller classes in Java).
Natural language-based techniques alleviate such biases and constitute the state-of-the-art in software modeling assistance.
\citet{arora2016extracting} propose an approach to automatically extract domain models from requirement documents by rule-based techniques. Their results show promising results with about $74-100\%$ recommendations deemed eventually correct in their experiments.
~\citet{burgueno2021nlp} exploit language models that are trained on general natural language documents to suggest domain concepts for model completion, and reach up to 62\% reliability in identifying model elements.
Unfortunately, natural language-based techniques provide solutions to a specific case and re-training is required for new cases, problems and projects. Such resource-intensive endeavors typically cannot be justified in practice, limiting the applicability of  powerful and efficient natural language-based techniques.
However, the recent emergence of large language models  has opened up new opportunities for using natural language to help with modeling.  
In a study by \citet{wang2024llms}, students interacted with LLMs and defined their own prompts to create use case models, class diagrams, and sequence diagrams. The results showed that while LLMs can provide valuable assistance, they also have notable shortcomings and limitations, such as difficulty in identifying and analyzing relationships and instability in producing fully compliant and correct UML models. In contrast, our work provides a systematic approach with predefined prompts and structured guidance, enhancing the usability and effectiveness of LLMs in domain modeling.

\subsection{Large Language Models and Few-shot Learning}\label{sec:background-llm}

Large language models (LLMs) are AI models pre-trained on vast amounts of textual data, enabling them to capture extensive human-generated knowledge and use this encoded knowledge for generative tasks such as creating human-like text~\cite{ozkaya2023application}. The key mechanism of LLMs relies on predicting the conditional probability of a sequence of tokens based on previously observed tokens, e.g., generating the next word in a text. It is clear that the often human-like qualities of LLMs' output come from training it over a lot of data and the statistical rules that human language tends to follow ~\cite{hindle2012naturalness}.
The GPT family\footnote{\url{https://openai.com/gpt-4}} comprises the largest and most adopted LLMs at the time of writing this article. It has been used in numerous tasks in the overall software development lifecycle, from requirements engineering~\cite{ronanki2023investigating} through software architecture~\cite{ahmad2023towards} to testing~\cite{guilherme2023initial}; as well as organizing software engineering knowledge by accelerating literature surveys~\cite{syriani2024screening,syriani2023assessing}.

In most occasions, before employing them, LLMs need to be fine-tuned. This is because of the generality of the corpus LLMs are pre-trained on. Fine-tuning adjusts model parameters slightly, rendering the LLM more adapted for a particular task. Thanks to the extensive nature of pre-training, fine-tuning is usually a minor effort that includes (i) hyperparameter tuning and (ii) instruction tuning ~\cite{zhang2023instruction}. Some of the main hyperparameters of a GPT model are \textit{temperature}, controlling the randomness, creativity and diversity of the response and \textit{maxTokens}, controlling the length of the response. We elaborate on our choices of hyperparameters in \secref{sec:tool}.
Instruction tuning is achieved through systematically engineering prompts, which are brief textual inputs that  LLMs are capable of responding to. By varying the wording of the prompt, different responses can be obtained~\cite{white2023prompt}.

Expected input-output pairs, known as \textit{shots}, may be included in certain prompt instances.  Depending on the number of shots, we distinguish between zero-shot learning (no shots are provided), single-shot learning (one shot is provided), and few-shot learning (a small number of shots are provided)~\cite{liu2023pre}.
%
We employ few-shot learning, where shots are generated from already existing segments of the model under construction. (See \secref{sec:approach} for details.)

Within software engineering, Model-Driven Software Engineering (MDSE) is in a prime position to leverage LLMs due to its emphasis on the formal semantics of languages ~\cite{schmidt2006guest}.  While LLMs can produce output that often exhibits human-like qualities, it is still not guaranteed that these outputs make sense~\cite{arkoudas2023chatgpt}. Therefore, a systematic approach to mapping LLMs' outputs to the target language's semantics is crucial for reliable LLMs-powered modeling methods in MDSE. As a consequence, there has been a surging interest in employing LLMs in MDSE lately, e.g., in automated software modeling~\cite{hou2023large,wang2024llms,chen2023automated}. In a related work, \citet{tinnes2024leveraging} achieves a semantic correctness of 62.30$\%$ via retrieval-augmented generation in a real-world industry setting.

Conversely, the benefits of MDSE \textit{for} LLMs have been recognized, e.g., in model-driven prompt engineering. For example, \citet{clariso2023model} propose using domain-specific languages (DSLs) to define platform-independent prompts, which can then be adapted to generate high-quality outputs in target AI systems. Unfortunately, no systematic approach exists for prompting LLMs for domain model completion that involves semantic mappings, efficient tooling, and a clear evaluation of utility. In the following sections, we present such an approach and reflect on its challenges.
\section{Prompting LLMs for Domain Model Completion}\label{sec:approach}

As highlighted by Camara et al. ~\cite{CamaraTBV23}, while LLMs excel in many tasks, they are not yet fully capable of automating model creation, especially for domain-specific languages. Nevertheless, LLMs have accumulated a lot of valuable knowledge that can be used in modeling activities such as model completion.
However, a crucial gap exists between inferring accurate knowledge for a task and effectively leveraging that knowledge to benefit users. This paper focuses on exploring the utility of automated domain model completion using LLMs. 
To evaluate this utility, we propose an approach and a tool specifically designed to assist modelers in their tasks. This section introduces our approach and provides a practical illustration using a running example. Our proposed approach builds upon our prior work on domain model completion utilizing LLMs~\cite{benchaaben2022towards}. Our approach is based on an iterative process with the domain modeler in the loop. Each iteration has a series of steps.
First, we read the model under construction (either the textual representation of graphical models or its textual serialization -- e.g. XMI, UML or Ecore models), and use parsers to automate the extraction of information. In particular, we extract information about each class name, its attributes and the associations between classes. 
Second, we establish what we refer to as \textit{semantic mappings}. A semantic mapping is defined as the translation of elements within the domain model under construction into semantically equivalent text. For this, we use the obtained information and filter it depending on the purpose (i.e., create class suggestions, attribute suggestions and association suggestions). 
Third, each piece of generated text in the previous step is integrated into a prompt, ensuring its comprehensibility to the LLM at a later stage. We create different prompts for different kind of suggestions (e.g., new classes, new attributes, new associations).
Fourth, we prompt the LLM, collect and parse the result and generate suggestions.
Suggestions are presented to the modeler who can accept or ignore them. If accepted, the suggestion is integrated in the model under construction.
These steps are presented with further details in Algorithm~\ref{alg:algoritm}.

\begin{algorithm}
\small 
\DontPrintSemicolon
\textbf{def} suggestionGeneration() \Begin{
    \While {model is not finished} {
        \textit{textualInfo} $\leftarrow$ readAndExtractInformation(model)\;

        \textit{classNames} $\leftarrow$ filter4Classes(textualInfo)\;
        \For {i = 1 to n} {
            \textit{gc} $\leftarrow$ shuffle(classNames)\;
            \textit{suggestionsC} $\leftarrow$ \textit{suggestionsC} + suggestions4Classes(gc)\;
        }
        rankSuggestions(suggestionsC)\;
        
        \textit{groupsForAttSugg} $\leftarrow$ filter4Attributes(textualInfo)\;
        \For {ga : groupsForAttSugg} {
            \textit{suggestionsAtt} $\leftarrow$ \textit{suggestionsAtt} + suggestions4Attributes(ga);
        }
        rankSuggestions(suggestionsAtt)\;
        
        \textit{groupsForAssocSugg} $\leftarrow$ filter4Associations(textualInfo)\;
        \For {gass : groupsForAssocSugg} {
            \textit{suggestionsAssoc} $\leftarrow$ \textit{suggestionsAssoc} + suggestions4Assoc(response, `assoc')\;
        }
        rankSuggestions(suggestionsAss)\;
        
        presentSuggestions(suggestionsC, suggestionsAtt, suggestionsAssoc)\;
        accepted $\leftarrow$ acceptSuggestion(suggestion)$^*$\;
        \If {accepted} {
            incorporateToModel(suggestion)\;
        }
    }
}
\textbf{def} suggestions4Classes(elements) \Begin{
    \textit{prompt} $\leftarrow$ createPrompt(elements, `class')\;
    \textit{response} $\leftarrow$ prompt(llm, prompt)\;
    \textit{suggestions} $\leftarrow$ parseAndBuildSuggestions(response, `class')\;
    \nl\KwRet{$suggestions$}
}
\textbf{def} suggestions4Attributes(elements) \Begin{
    \textit{prompt} $\leftarrow$ createPrompt(elements, `att')\;
    \textit{response} $\leftarrow$ prompt(llm, prompt)\;
    \textit{newAttributeSugg} $\leftarrow$ parseAndObtainNewAttributeSugg(response, `att')\;
    \For {a : newAttributeSugg} {
        \textit{prompt'} $\leftarrow$ createPrompt(a, `attType')\;
        \textit{response} $\leftarrow$ prompt(llm, prompt')\;
        \textit{typeSugg} $\leftarrow$ parseAndObtainNewAttributeTypeSugg(response, `attType')\;
        \textit{suggestions} $\leftarrow$ buildAttributeSuggestion(a, typeSugg)\;
    }
    \nl\KwRet{$suggestions$}
}
\textbf{def} suggestions4Assoc(elements) \Begin{
    \textit{prompt} $\leftarrow$ createPrompt(elements, `assocNames')\;
    \textit{response} $\leftarrow$ prompt(llm, prompt)\;
    \textit{newAssocNames} $\leftarrow$ parseAndObtainNewAssocSuggestions(response, `assocNames')\;
    \For {a : newAssocNames} {
        \textit{prompt'} $\leftarrow$ createPrompt(a, `assocType')\;
        \textit{response} $\leftarrow$ prompt(llm, prompt')\;
        \textit{typeSugg} $\leftarrow$ parseAndObtainNewAssocTypeSugg(response, `assocType')\;
        \If {typeSugg = `Inheritance'} {
            \textit{prompt''} $\leftarrow$ createPrompt(a, `inhDirection')\;
            \textit{response} $\leftarrow$ prompt(llm, prompt'')\;
            \textit{directionSugg} $\leftarrow$ parseAndObtainNewAssocTypeSugg(response, `inhDirection')\;
        }
        \textit{suggestions} $\leftarrow$ buildAssocSuggestion(a, typeSugg, directionSugg)\;
    }
    \nl\KwRet{$suggestions$}
}

\caption{Iterative process for model completion. }
\label{alg:algoritm}
\end{algorithm}

As mentioned before, depending on the nature of the suggestions that we would like to provide, 
we need to adapt our semantic mappings. In the following, we will provide all the details and we illustrate them using our running example.





\begin{figure}[h]
\centering
\includegraphics[width=0.6\columnwidth]{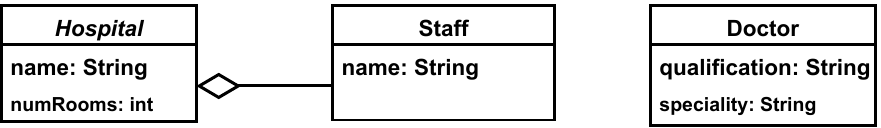}
\caption{Hospital domain model (in construction)}
\Description{Invalid}
\label{fig:runningExample}
\end{figure}

Let us assume that a modeler is building a domain model to capture information about hospitals. At a certain point in time, the model is in the state that ~\figref{fig:runningExample} captures. It has only 3 basic entities: \texttt{Hospital}, \texttt{Staff} and \texttt{Doctor}. A hospital has an attribute of type String named \texttt{name}, an attribute \texttt{numRooms} of type int and one aggregation to the entity \texttt{Staff}. \texttt{Staff} has one attribute \texttt{name}, and \texttt{Doctor} has two  attributes \texttt{speciality} and \texttt{qualification}.

Regarding the creation of prompts, we incorporate the principles in \cite{PromptingGuide2023,liu2022design} to meticulously craft prompts tailored to our objectives. We use the strategy few-shot prompt learning. Therefore, each prompt  is composed of three critical elements: the \emph{Instruction} outlining the task at hand, \emph{Few-shot examples} providing context and guidance, and the \emph{encoded Model under construction}, all aimed at facilitating the completion of the model.

\subsection{Suggesting model elements}

\subsubsection{Suggesting classes}

For class suggestion, the goal is to obtain meaningful names for classes to recommend. The instruction within the prompt is the text `\textit{Generate related concepts}'. 
The list of shots (examples) is extracted from a catalog of diagrams from diverse and unrelated domains.
For each shot $i$, first, we add to our prompt the package name ($X_i$) to establish the context of the domain, then we add pairs of classes (i.e., $Y_{ij}$ and $Z_{ij}$) that are connected with a relationship.
We use the following syntax: $X_i: [Y_{i1},Z_{i1}], [Y_{i2},Z_{i2}], .... [Y_{in},Z_{in}]$.
Finally, to encode the model under construction, we use the elements extracted from the domain model for class suggestion -- which are the class names -- and put them in square brackets, separated by commas. 
We append to the prompt this information using the same syntax as for the few shots.


Listing~\ref{lst:promptClasses} shows an example of a prompt to suggest new classes for our running example where we have used two diagrams (\textit{Banking System} and \textit{Reservation System}) to build the few-shots.

\begin{lstlisting}[numbers=none,caption={Example of prompt for class suggestion generation},captionpos=b,label=lst:promptClasses]
    Generate related concepts:
    BankSystem: [Bank, ClientCollection], [ClientCollection, Client]
    ReservationSystem: [Airport, City], [Trip, Passenger], [Passenger, Plane]
    HospitalSystem: [Hospital, Staff], [Doctor, Staff]
\end{lstlisting}

Once the LLMs is prompted and the response is available, we need to parse it and build the suggestions. In our example, we have prompted GPT-3. The response is the text displayed in Listing~\ref{lst:responseClasses}. 
We apply a string-searching algorithm to the generated text to extract relevant class names and the association that exists between them. Next, we put in place a post-processing phase in which we remove spelling errors and noisy data such as digits, which are usually not part of domain models. Then, we obtain the potential missing classes (and associations between them). In our example, these are \texttt{Patient}, \texttt{Appointment} and \texttt{Address}. These suggested classes are then presented to the modeler after ensuring they are not already present within the canvas.

\begin{lstlisting}[numbers=none,caption=Completed text by GPT-3 for class suggestion generation,captionpos=b,label=lst:responseClasses]
    [Patient, Appointment], [Address, Hospital]
\end{lstlisting}


Our engine implements a ranking strategy when suggesting new elements. We create different prompts and query the LLM several times (see the for loop in line 5 of Algorithm~\ref{alg:algoritm}). All prompts share the same few shots but different queries. Each query contains the same subset of elements from the partial model but in different orders. As a result, we obtain, for each prompt, a set of suggested concepts. Then, all the obtained concepts from the different prompts are ranked by their frequency, from higher to lower. Only those concepts with higher frequency are suggested to the user.

\subsubsection{Suggesting attributes}


To generate attribute suggestions for a class, we follow a two-step process. First, we obtain attribute names, and then, for each name, we extract the attribute type.

During the first step, the instruction within the prompt is ``Generate missing attributes for each class in this class diagram''. The list of shots is again extracted from the previously mentioned catalog of diagrams. 

For each shot, we gather a set of classes from a package $i$. We then combine the package name ($X_i$) with each class name $Y_{ij}$ and its attributes $A_{ij1}..A_{ijn}$ in square brackets. Following this, we append the symbol "=>" and merge another set of classes with their attributes.

That is, we employ the following syntax: $X_i: Y_{i1}: [A_{i11},A_{i12},...A_{i1n}]; ..., Y_{ik}: [A_{ik1},A_{ik2},...A_{ikn}]]$.
Finally, to encode the model under construction, we use the elements extracted from the domain model for attribute suggestion (i.e., \textit{ga} in Algorithm \ref{alg:algoritm}). We concatenate the package name, the existing class names with their attributes in square brackets, the class for which we are finding potential attributes and the symbol ``=>''.

In our running example, a potential prompt (with only one shot) is shown in Listing ~\ref{lst:promptAttributes}.

\begin{lstlisting}[numbers=none,caption=Prompt for attribute suggestion generation,captionpos=b,label=lst:promptAttributes]
Generate missing attributes for each class in this class diagram:
package company: employee: [id, name, lastName, occupation]; manager:[id, name, department];
company: [name, holding] => employee: [id, name,lastName]; manager: [id, name, department, team, revenue];
company: [name, holding, address, website]
package Hospital: Hospital: [name, rooms number]; Staff: [name]; Doctor:[speciality,qualification]; Patient[]; Appointment:[]; Address: [] =>
\end{lstlisting}

After prompting GPT-3, we obtain what is presented in the Listing~\ref{lst:responseAttributes}. We have highlighted in bold the new attributes that the LLM is suggesting. Like before, we need to parse the response to extract the new suggestions. 

\begin{lstlisting}[numbers=none,caption=Response for attribute suggestion generation,captionpos=b,label=lst:responseAttributes,mathescape=true]
Hospital: [name]; Staff: [name, speciality, $\textbf{salary}$]; Doctor: [speciality, qualification]; Patient: [name, $\textbf{id, phone number}$]; Appointment: [$\textbf{date, time, doctorName}$]; Address: [$\textbf{street, city, state, postal code, country}$]
\end{lstlisting}

Next, for each attribute, we need to obtain its type, for which we prompt the language model again, with the description ``Generate attribute type''. In the prompt, as shots, we provide a pseudo-random set of attribute names $A_i$,  where this set contains at least one attribute of each datatype, along with their associated types $T_i$ from our catalog, followed by the attribute name whose type we want to predict. We use the syntax: $A_i => T_i$ for each pair and separate them with commas. 
The LLM generates the most suitable type for the specified attribute in its response. An example is illustrated in Listing~\ref{lst:promptAttributeType}. When GPT-3 is prompted, it returns the attribute's type \textit{String} (see Listing~\ref{lst:responseAttributeType}).

\begin{lstlisting}[numbers=none,caption=Prompt for attribute type generation,captionpos=b,label=lst:promptAttributeType,mathescape=true]
Generate attribute type: 
Address => String, age => int, name => String, isCanceled => boolean, sold => float, surname => String, birthDate => Date, isValidated => boolean, staffNumber => int, width => double, phoneNumber => float, city => String, state => String, street =>
\end{lstlisting}

\begin{lstlisting}[numbers=none,caption=Response
for attribute type generation,captionpos=b,label=lst:responseAttributeType,mathescape=true]
street => $\textbf{String}$
\end{lstlisting}


Finally, we use a frequency-based ranking function that takes as input all the attributes generated by the LLMs for the different prompts and we present to the modeler only those at the top of the ranking.

\subsubsection{Suggesting associations}
\label{subsubection:Associations}

To generate association suggestions for a domain model, we follow three steps. First, we try to identify whether there is a relationship between two classes by obtaining the association's name. Second, we try to obtain the kind of association. Finally, if inheritance is identified, we determine its direction.

To obtain association names, we use the following instruction that is included in the prompt ``Predict association name''. The list of shots is again compiled from the catalog of diagrams, from which we select random pairs of classes that have an association between them. For each pair, we concatenate the name of the classes $C_i, C_j$ and the name of the association $A_{ij}$ using the following syntax: $C_i, C_j => A_{ij}$. Each group of elements is separated with a semicolon. Then, we take the model under construction, and for each association, we generate the text using the same syntax and add it to the prompt. An example can be seen in Listing~\ref{lst:promptAssociation}. Note that this prompt only contains \emph{Instruction} and \emph{Shots}. As there are no associations with name in the partial model, the partial model is not encoded nor added to the prompt.

\begin{lstlisting}[numbers=none,caption=Prompt for association suggestion generation,captionpos=b,label=lst:promptAssociation,mathescape=true]
Predict association name:  
employee, company => worksIn ; person, Home =>  owns ;  car,driver => drives  ; personalCustomer, customer => is ;  man, women=>  married ; lion, meat => eats ; manager, employee  =>  supervises ; order,  customer =>   places ; user, account =>  has ; cat, milk => drinks ; employee, department =>  worksIn ;
\end{lstlisting}


To generate suggestions for the type of associations, the instruction included in the prompt is ``Specify the nature of the association between these concepts, inheritance or association or composition or no''. Each pair of classes $C_i, C_j$ is used to construct a shot by concatenating the class names and the association type $A_{ij} \in \{$inheritance, association, composition, no$\}$ as follows: $C_i, C_j => A_{ij}$. Then, the two class names for which the type of association wants to be predicted are appended to the prompt followed by ``=>''. Listing~\ref{lst:promptAssociationType} presents a prompt for our running example. When prompting GPT-3, the response is: ``inheritance''.

\begin{lstlisting}[numbers=none,caption=Prompt for association type suggestion generation,captionpos=b,label=lst:promptAssociationType,mathescape=true]
Specify the nature of the association between these concepts: inheritance or association or composition or no:
Student, Person => inheritance \n Computer, Cpu => composition \n Plane, Passenger => no \n Person, Address => association \n
Doctor, Staff =>
\end{lstlisting}

Finally, when the relationship predicted between two classes is inheritance, we prompt the LLMs once again  to determine which class is the superclass. The instruction of this prompt is: ``Select the  super class in this UML inheritance relationship''. The shots are pairs of subclasses and superclasses from our catalog $C_i, C_j$ followed by ``=>'' and the superclass ($C_i $ iff $C_i > C_j$ or $C_j $ iff $C_j > C_i$). Then we append the two classes for which we want to predict the direction of the inheritance. Listing~\ref{lst:promptInheritanceDirection} shows an example of the prompt.

\begin{lstlisting}[numbers=none,caption=Prompt for association suggestion generation,captionpos=b,label=lst:promptInheritanceDirection,mathescape=true]
Select the  super class  in this UML inheritance relationship:
admin, user  => user \n  Account , SavingAccount => Account \n Room, doubleRoom  =>Room \n vehicle, car => vehicle \n dog, animal => animal \n 
staff, doctor =>
\end{lstlisting}

\begin{figure}[t]
  \centering
  \includegraphics[width=\columnwidth]{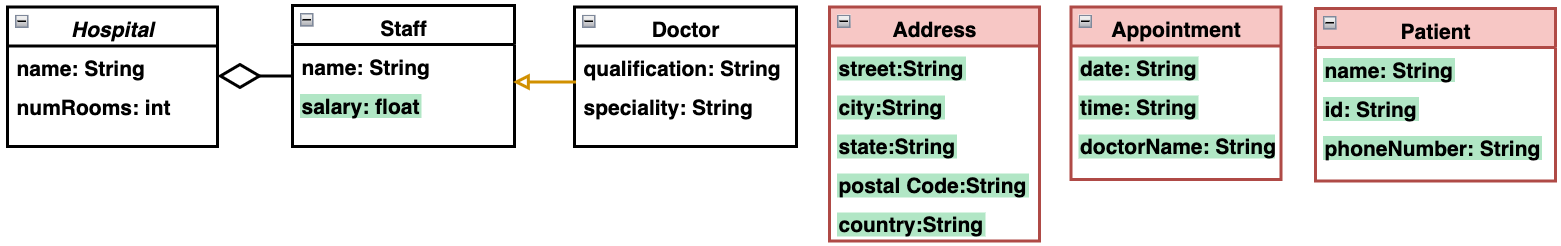}
  \caption{Completion Suggestions for Hospital Domain Model (classes in red, attributes in green, association in yellow).}
  \Description{Invalid}
  \label{fig:completion}
\end{figure}

~\figref{fig:completion} presents the suggestions that have been generated by our approach for the model under construction of ~\ref{fig:runningExample}.

\subsection{Suggestion modes}

To present the suggestions for the user, we implement three modes that address different stages of the modeling process and user preferences.

\subsubsection{Suggestions On request}
This mode allows users to actively seek out specific predictions at any point in the modeling process, aiming for precise searches of attributes, concepts, or associations. This flexibility is particularly useful for users who have a clear understanding of their needs and prefer to control when and what assistance they receive. The rationale behind this mode is based on the principles of direct manipulation interfaces, which support user control, as discussed in ~\cite{shneiderman1997direct}.
 
\subsubsection{Automatic suggestions} This mode provides real-time, automatic suggestions of relevant concepts and operations as users add elements to their model. It offers continuous support that adapts to the evolving context of the model. This mode helps maintain a steady flow of relevant ideas and ensures that the assistance is proactive and contextually aware ~\cite{dey2001understanding}. Additionally, addressing interaction challenges with automated systems, as highlighted by ~\citet{wessel2021don}, ensures that suggestions are helpful without being disruptive.

\subsubsection{Suggestions at the end}
 This mode presents suggestions for potential enhancements upon model completion, enabling users to conduct a final review and  refinement. This mode is particularly useful for users who prefer to first develop their initial ideas independently and then seek out ways to refine and enhance their work based on comprehensive feedback. Such an approach aligns with the principles discussed in ~\cite{cross2021engineering} which emphasizes the importance of post-design evaluation and refinement to enhance the quality and functionality of the final product.

\section{Tool support}\label{sec:tool}

Conducting our experiments requires a suitable modeling tool. Unfortunately, state-of-the-art tools do not offer proper integration capabilities with LLM services or flexible customization of their internals. Therefore, we developed \textbf{MAGDA} (\textbf{M}odeling \textbf{A}ssistance with \textbf{G}enerative Ai for \textbf{D}omain represent\textbf{A}tion), a research tool to support our work\footnote{The tool is available as an open-source project: \url{https://github.com/geodes-sms/MAGDA}.}.
When creating the tool, we aimed to reduce potential validity threats from inadequate instrumentation or tools.

\begin{figure}[H]
  \centering
  \includegraphics[width=0.7\columnwidth]{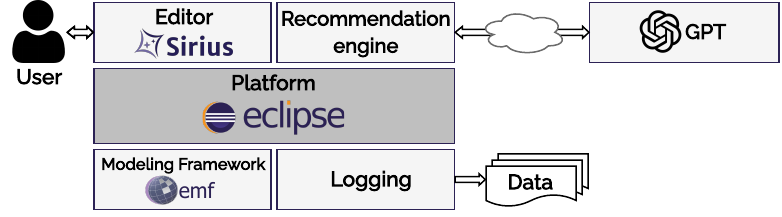}
  \caption{High-level architectural view of MAGDA}
    \Description{Invalid}

  \label{fig:tool}
\end{figure}

The high-level overview of the tool is shown in \figref{fig:tool}.
We chose the \textit{Eclipse platform}\footnote{\url{https://www.eclipse.org/}} to implement our tool thanks to the rapid development curve it enables through its ready-to-use modeling framework and extensible plug-in--based architecture.
The \textit{user} interacts with the models through the visual \textit{Editor} (\secref{sec:tool-editor}) that allows the creation of domain models using UML Class Diagrams.
The \textit{Modeling Framework} (\secref{sec:tool-modeling}) serves for representing and persisting models.
The \textit{Recommendation engine} (\secref{sec:tool-recoengine}) is responsible for querying the LLM ( \textit{GPT}  Model) upon request or upon edit operations; collecting, organizing, and ranking recommendations; and passing this information to the editor by saving recommendations in the model.
Finally, to investigate user behavior during modeling, we developed a \textit{Logging} facility~(\secref{sec:tool-logging}).

In the following, we provide further details about these technical components.

\subsection{Editor}\label{sec:tool-editor}

Modeling is supported by a visual editor that we have developed using the Eclipse Sirius language workbench\footnote{\url{https://eclipse.dev/sirius/}}. Building a modeling environment with Sirius is achieved in three steps. First, we define a domain model for class diagrams and recommendations (as shown and discussed in detail in \figref{fig:metamodel}). Second, we define the concrete syntax, i.e., the notation the user interacts with when defining class diagrams. We follow the standard UML notation. Third, we develop productivity features to improve the user experience and mitigate any threats to the construct validity of our study that might stem from inappropriate tooling.

\begin{figure}[ht]
    \begin{center}
    \begin{mdframed}[backgroundcolor=white]
        \includegraphics[width=1\textwidth]{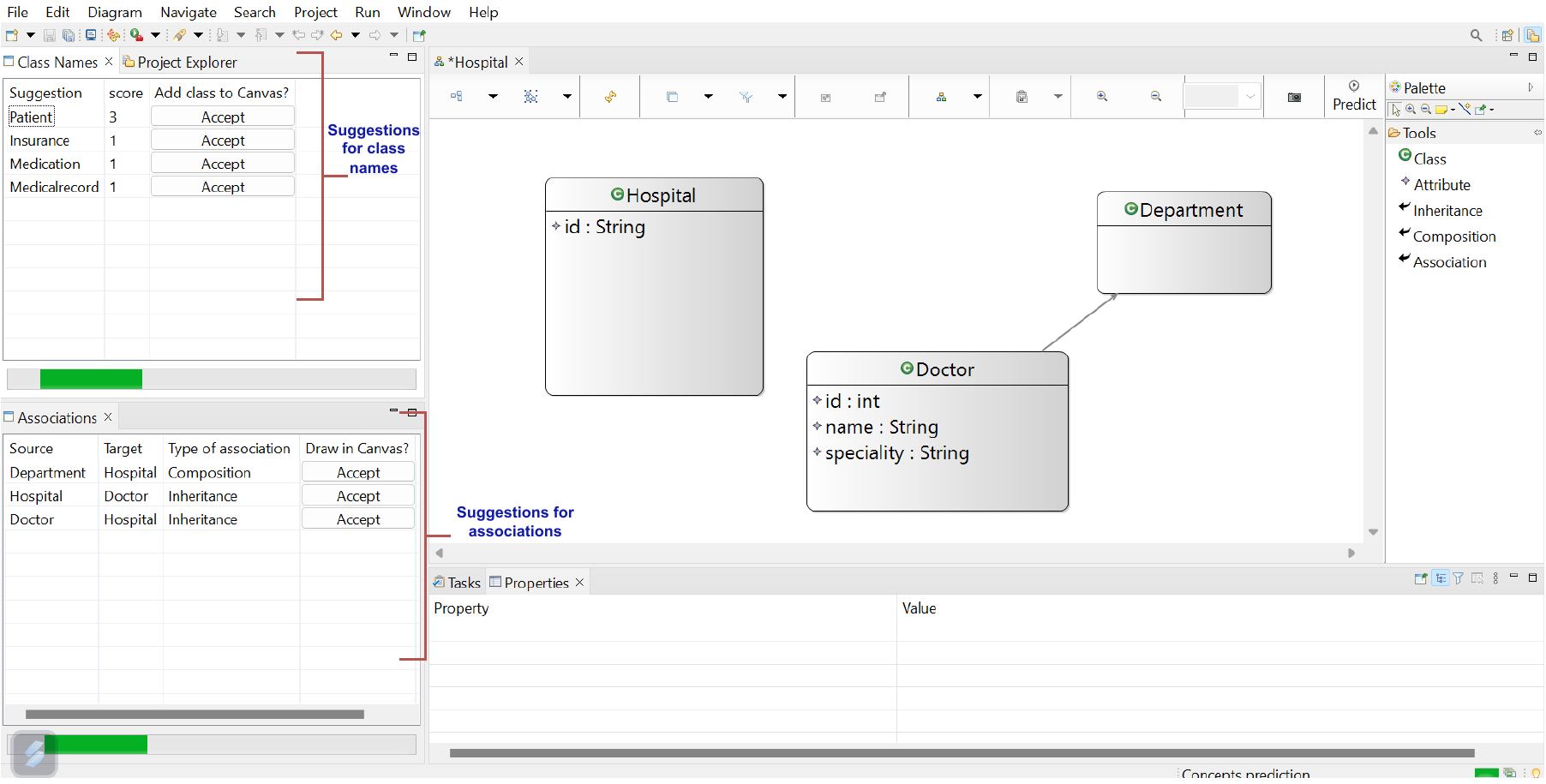}
    \end{mdframed}
    \caption{User interface of the model editor}
      \Description{Invalid}
    \label{fig:tool-ui}
    \end{center}
\end{figure}

The eventual user interface is shown in \figref{fig:tool-ui}.
In the middle of the screen, a canvas shows the elements of the constructed class diagram. On the right side, a toolbox allows the user to edit the model on the canvas. On the left side, the recommendation panels show the list of recommended classes (top left) and the list of recommended associations (bottom left). Next to each recommendation, an action button that allows the user to \textit{accept} the recommendation, which automatically places the element on the canvas.
This step is better explained in the next section, where we elaborate on the relationship between the actual class diagram and recommended class diagram elements.

\subsection{Modeling framework}\label{sec:tool-modeling}

\begin{figure}[ht]
  \centering
  \includegraphics[width=0.6\columnwidth]{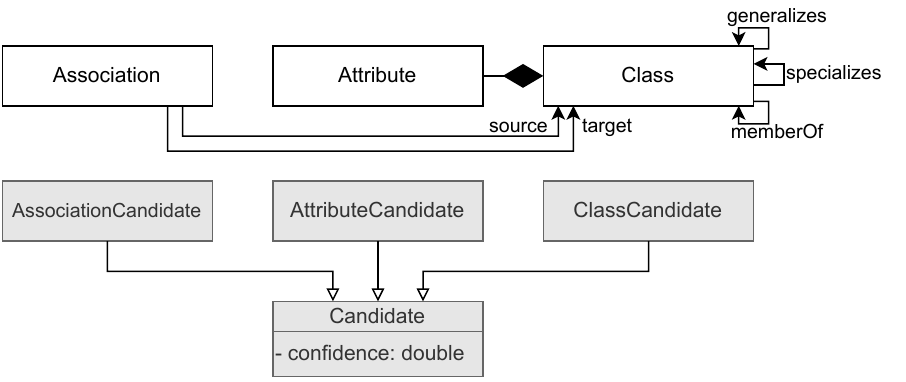}
  \caption{Simplified excerpt from the metamodel of the modeling framework of MAGDA}
    \Description{Invalid}

  \label{fig:metamodel}
\end{figure}

The modeling framework serves to represent and persist models. We used the Eclipse Modeling Framework (EMF) for this purpose\footnote{\url{https://eclipse.dev/modeling/emf/}}. EMF is a collection of Eclipse plug-ins allowing for modeling a domain and generating code (e.g., EDIT transaction handlers and user interface components).

We defined the metamodel shown in \figref{fig:metamodel} that allows for representing class diagrams (white components), as well as the \textit{candidate} counterparts of class diagram elements (gray components).
The user captures the model in terms of \textit{Class}es, \textit{Attribute}s, and \textit{Association}s between \textit{Class}es. Each of these metamodel elements has a candidate part---i.e., \textit{ClassCandidate}, \textit{AttributeCandidate}, and \textit{AssociationCandidate}, respectively---to represent recommendations by the LLM. Once a recommendation is accepted by the user, the corresponding candidate element is removed from the model and a corresponding real model element is created. For example, upon accepting a recommended \textit{ClassCandidate}, a model element of \textit{Class} is created.
Additionally, every model element has a name and type they inherit from the superclasses NamedElement and TypedElement, respectively (Not shown in the figure, for the sake of simplicity.)

Candidate model elements are equipped with a \textit{confidence} attribute we use to calculate the level of trust in the given recommendation. We calculate this value based on the number of times the element has been recommended by the LLM during the modeling session. The more frequently an element is suggested, the higher its confidence value. 

We use this data to organize recommendations in descending order of confidence and present the top twenty recommendations if there is an excessive number. Confidence values are transparently provided to users to allow for informed decision-making.

\subsection{Recommendation engine}\label{sec:tool-recoengine}

The recommendation engine is responsible for interacting with the LLM to generate, collect, collate, and persist recommendations by the metamodel in \figref{fig:metamodel}.

\subsubsection*{Using GPT as the LLM}
The recommendation engine in our experiments uses GPT-3\footnote{\url{https://openai.com/index/gpt-3-apps}. }. A GPT model is an appropriate choice for our purposes as the most powerful LLM at the time of the experiments and at the time of writing this article. We integrate MAGDA with GPT through the OpenAI's API\footnote{\url{https://platform.openai.com/docs/api-reference}}. We parameterize GPT in accordance with community practices. Choosing tested and proven settings mitigates threats to validity.
The key parameters are listed in \tabref{tab:gpt-params}.

\begin{table}[H]
\centering
\caption{Parameters used to configure GPT in our experiments}
\small
\begin{tabular}{@{}llrl@{}}
\toprule
\multicolumn{1}{c}{\textbf{Parameter}} & & \multicolumn{1}{c}{\textbf{Value}} &
\multicolumn{1}{c}{\textbf{Reason for choice}}\\ \midrule
model && text-davinci-002 & Advanced language understanding~\cite{zhao2023survey} \\ \hline
\code{temperature} && 0.7 & Balance between creativity and consistency~\cite{mediumGPT3Temperature} \\\hline
\multirow{4}{*}{\code{maxTokens}} & Class & 8 & \multirow{4}{*}{Reasonably short and concise responses~\cite{openai2024what}} \\
& Attribute & 40 & \\
& Attribute type & 2 & \\
& Association (type) & 2 & \\
\bottomrule
\end{tabular}
\label{tab:gpt-params}
\end{table}

We opt for the \textit{text-davinci-002} engine\footnote{\url{https://platform.openai.com/docs/models/gpt-base}} due to its superior ability to generate English text.
We also set two important hyperparameters: \code{temperature} and \code{max\_tokens}.
The \code{temperature} parameter controls the randomness of the output generated by GPT. Higher temperatures result in more variance in the output, which can be interpreted as higher ``creativity''. In our experiments, we require a fine balance between creativity and consistency, and therefore, set the \code{temperature} to 0.7, which is considered the lower threshold for creative tasks~\cite{mediumGPT3Temperature}.
The \code{max\_tokens} parameter controls the length of the generated text. One token roughly equals to 4 characters or \sfrac{3}{4} words in English~\cite{openai2024what}. We aim to keep the responses reasonably short and concise.
For class recommendations, we set \code{max\_tokens} to 8, which is approximately equal to 32 characters or six words. Given the naming conventions in software engineering, specifically that concepts sometimes tend to be compound words, these six words correspond to about four-to-six recommended concepts.
For full attribute recommendations, we set \code{max\_tokens} to 40, which is approximately equal to 160 characters; and for attribute types, we set the parameter to 2. At this level, the typical naming conventions of software engineering take a strong bias towards long names. Reportedly, professionals find longer names with more embedded sub-words more informative~\cite{liblit2006cognitive} and more insightful~\cite{lawrie2006what}. Based on our pilot experiments, we estimate that a typical full attribute recommendation (name and type) would be around 4--7 tokens, allowing for about 6--10 attributes to be recommended upon each query.
Finally, for associations, we use two tokens to represent both information: the type and the name. 

\subsubsection*{User experience (UX) considerations} We took a number of measures to ensure a smooth modeling experience.
We developed our tool as a multi-threaded application in which users can interact with the model while recommendations are fetched from the LLM asynchronously, without blocking the graphical user interface (GUI). We rely on the Jobs API of Eclipse\footnote{\url{https://help.eclipse.org/latest/topic/org.eclipse.platform.doc.isv/reference/api/org/eclipse/core/runtime/jobs/Job.html}} to develop multi-threading. Jobs are used to query GPT, populate the recommendation panels, and handle user interactions. Recommendation panels are updated as soon as new recommendations are retrieved. This allows for improved productivity and an intuitive modeling experience.

To improve the efficiency of working with a remote LLM such as GPT, we implemented a cache system.
The cache system is responsible for storing and retrieving frequently used data and avoiding redundant calls to GPT. This allows us to reduce the delivery time of recommendations substantially.

Additionally, we implemented a backoff mechanism for API calls. Backoff is the technique of increasing the time between subsequent API calls in case of a failing or thrashing remote system\footnote{\url{https://platform.openai.com/docs/guides/rate-limits/error-mitigation}}. Backoffs helps us manage the volume of requests toward GPT and by that, maintain system stability. Backoff further prevents overloading the remote service (here: GPT-3), ensuring reliable interactions. To enhance recommendation relevance, we query the GPT model multiple times for each request and present the user with frequently recommended elements. This method ensures more accurate and contextually relevant suggestions, improving the overall modeling experience and precision.

\subsection{Logging}\label{sec:tool-logging}

To analyze participants' behavior, we collect data about their modeling exercise. To this end, we developed a logging system that records (i) every user operation that affects the model, such as creating and deleting model elements, and accepting recommendations; (ii) user requests for recommendations; and (iii) the recommendations generated by the LLM. Records are timestamped, allowing us to unambiguously reconstruct the modeling exercise from the logged data.
Logs are persisted on the file system for further investigation in comma-separated files. The replication package contains the log files of the experiments reported in this paper.\footnote{\label{note1}\url{https://github.com/meriembenchaaben/UtilityOfDomainModelCompletion-replicationPackage} -- The final version of the replication package is to be published on Zenodo after considering the feedback of reviewers.}
\lstref{lst:log-sample} shows a sample of the log files.

\begin{lstlisting}[numbers=left,caption=Excerpt from the log files,captionpos=b,label=lst:log-sample,]
timestamp, mode, operation, classes-real, class-reco, attrib-real, attrib-reco, assoc-real,assoc-reco
2023-05-08 12:00:43, auto, accept-view, {Student}, "School, Teacher", Student:[], School:[], "Teacher:[], Student:[]", , 
2023-05-08 12:01:02, auto, create-attribute, "School, Teacher, Student", "Course", "School:[], Teacher:[], Student[address]", "Student: [firstName, lastName, placeOfBirth, address, gender, level, middleName, dateOfBirth], Course:[]", , [student-school]=> ass, [teacher-school]=>ass
\end{lstlisting}

Model changes, recorded in the \texttt{operation} field of the records, including manual edits and acceptance of recommendations, are collected through the notification framework of EMF.\footnote{\url{https://download.eclipse.org/modeling/emf/emf/javadoc/2.11/org/eclipse/emf/common/notify/Notification.html}} EMF allows developers to get notified about changes in the model through defining observers and listeners for specific model elements or the whole model.
\section{Study Design}
\label{sec:studydesign}

In this section, we present the user study setup.
The complete protocol, materials, task descriptions, questionnaires, as well as the anonymized data we collected, are available as part of a replication package.\footref{note1}

\subsection{Research questions}
As outlined in the introductory section, we address the following five research questions. The first three pertain to the objective utility, while the remaining two questions focus on the utility as perceived by the modelers.

\begin{description}
    \item[RQ1 (Productivity).] \textit{What is the impact of modeling assistance on the \textbf{time} required to complete the domain models?}
    \item[RQ2 (Contributivity).] \textit{What is the level of contribution of the suggestions to the resulting models?}

    \item[RQ3 (Creativity).] \textit{Does the assistance reduce the solution \textbf{diversity} among the modelers?}

    \item[RQ4 (Assistance Preference).] \textit{What kind of assistance do participants \textbf{prefer to choose} when completing their tasks?}
    \item[RQ5 (Modeling experience).] \textit{How does the assistance affect the \textbf{modeling experience}, as perceived by the modelers?}

\end{description}

\subsection{Study locations and participant selection}

As part of our study, we have conducted a series of experiments (\secref{sec:experiment-setup}) in two different regions to improve the diversity of the study: at the University of Montreal in Montreal, Canada; and at the University of Malaga in Malaga, Spain; with 15 participants at each location, i.e., 30 participants total.


The Montreal experiments are carried out between May 1 -- May 10, 2023, conducted by the first author. 
The Malaga experiments are carried out between July 13 -- September 28, 2023, conducted by the second author. 
The language of experiments is English. We ensure that participants possess an appropriate command in English. In exceptional cases, the participant's native language may be used for clarification purposes between the researcher and the participant. 

To ensure consistent execution, we establish a protocol and adhere to it for every experiment. To validate the compliance with the protocol, the two researchers conducting the experiments randomly review each other's work and virtually attend and observe the procedures.


We did not face incidents during the experiments in Montreal. However, we experienced three incidents during the experiments in Malaga, in which unexpected technical problems surfaced leading to the experiments being terminated. We discarded those three incomplete experiments from our results and conducted three additional experiments (with new participants), resulting in 15 successful experiments at the Malaga location as well.

\begin{figure}[ht]
    \centering
    \begin{subfigure}[b]{0.32\textwidth}
        \centering
        \includegraphics[width=0.7\textwidth]{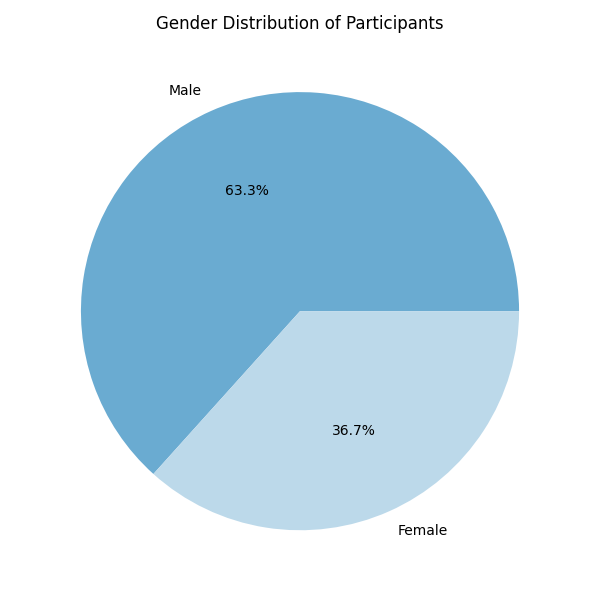}
        \caption{Gender distribution}
        \label{fig:gender_distribution}
    \end{subfigure}
    \hfill 
    \begin{subfigure}[b]{0.32\textwidth}
        \centering
        \includegraphics[width=\textwidth]{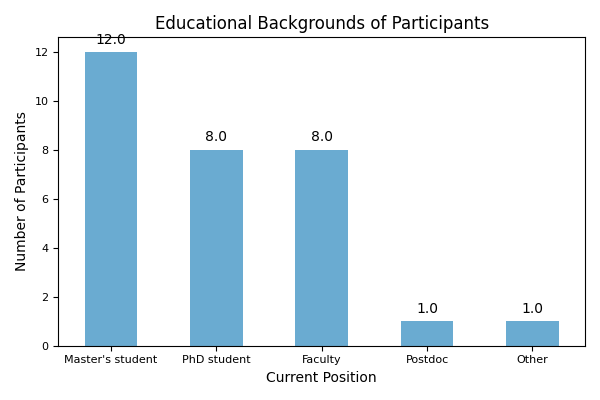}
        \caption{Educational background}
        \label{fig:educational_backgrounds}
    \end{subfigure}
    \hfill 
    \begin{subfigure}[b]{0.32\textwidth}
        \centering
        \includegraphics[width=\textwidth]{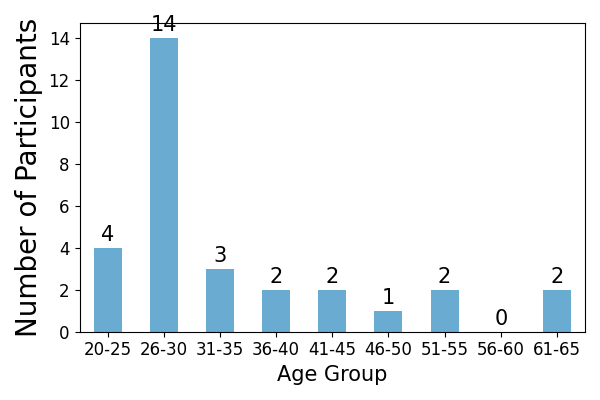}
        \caption{Age distribution}
        \label{fig:age_distribution}
    \end{subfigure}
        \caption{Participant demographics: (a) gender distribution with a male majority, (b) educational backgrounds dominated by Master's and PhD students, and (c) age distribution primarily in the 20-25 range.%
        }

    \label{fig:participant_demographics}
\end{figure}

We use an extended convenience sampling, mostly recruiting participants from the universities we are affiliated with (University of Montreal, Canada; and University of Malaga, Spain), as well as from universities in close geographical proximity in Montreal, Canada. To battle selection bias, we also ensure that the experience and level of expertise of the population is diverse enough. To this end, we invited undergraduate and graduate (Master's and PhD) students, postdoctoral researchers, professors, and industry experts as shown in ~\figref{fig:participant_demographics}. To mitigate threats to construct validity, we eventually recruit a potential participant if they possess an acceptable level of knowledge in domain modeling with UML Class Diagrams.

\subsection{Experimental setup and tasks}
\label{sec:experiment-setup}

Experiments are conducted in person, in an appropriately set up environment.

\paragraph{Pilot of the instrumentation}
Before the experiments, we run a pilot with one participant testing the setup, including the tool, the room, and the understandability of the problem descriptions. The pilot was concluded successfully, and its results were discarded.  

\paragraph{Preparation of participants}
Before an experiment takes place, each participant is provided with essential information about the basic workings of the tool used in the experiment. This information is shared the day before the experiment, in the form of a five-minute video, available online\footnote{\url{https://www.youtube.com/watch?v=5oUJJ0ial50}} and in the replication package\footref{note1}. At the beginning of the experiment, the conductor requests confirmation from the participant regarding the video watching and offers to address any questions that may have arisen.
\paragraph{Environment}

Experiments are conducted in a room with only a researcher and a participant present.
The participant is given access to a computer with the tool already installed and running. We use a standard computer system running a Windows operating system and equipped with a wide-screen monitor with a resolution of 1920$\times$1080.

\paragraph{Modeling}

Each experiment consists of completing four tasks within a one-hour time frame as follows. 

\begin{description}
    \item[Introduction.] At the beginning of the experiment, the lead researcher explains the details of the experiment
    to the participant and answers any questions the participant may have.

    \item[Tasks 1--3: Modeling based on a described problem.] The first three tasks require modeling a domain based on a short description of three-to-four paragraphs.
    
    We describe three domains: a hotel booking system (\textit{D1}), an online shop system (\textit{D2}), and a banking system (\textit{D3}). Modeling these domains requires a diverse set of modeling elements, thereby providing a sufficiently complete coverage of the syntax of the modeling formalism (UML Class Diagrams).
    
    For each task, the participant must use one of the recommendation modes (Automatic, On-Request, At-End). 
    The recommendation mode is pre-determined in the study design, as shown in \tabref{tab:participant_distribution}.
    As a baseline for comparison, we will also consider the no-assistance option. For the At-End task, participants are instructed to fully complete their models before receiving any suggestions. These pre-suggestion models will serve as the basis for comparison with the no-assistance condition, rather than creating a separate task for it.

    To mitigate threats to validity (learning and fatigue effects), we vary the distribution of recommendation modes in Tasks 1--3. We organize participants into three groups, and each group executes Tasks 1--2--3 using a different recommendation mode. \tabref{tab:task_order} summarizes the recommendation modes by experimental group and domain.
    
    For each task, participants are allocated 10 minutes, as a reference time. However, those who reach the 10-minute mark can be allowed some extra time to finish the part of the model they are working on. 
    
    After each task, participants are asked to fill in an online questionnaire about their modeling experience. The questionnaire contains a set of Likert-style items about positive and negative experiences that participants must answer on a 1--5 scale; and three open-ended questions regarding the features of the tool they missed, suggestions for improving the tool, and any other remarks. An example of a question about positive experiences is the following.``During the experiment, I felt \textit{confident}'' -- 1: less confident, 5: more confident. The questionnaire is available in \appref{app:questionnaire}. Detailed responses are available in the replication package\footref{note1}.
    
    \item[Task 4: Free modeling.] Participant are allowed to choose a domain of their preference, as well as any combination of recommendation modes. In this task, participants are allowed to switch between recommendation modes at any time.
    For this task as well, we allocate 10 minutes to complete.
    
    After this task, the participants fill out an exit questionnaire about their demographic data.
\end{description}

\begin{minipage}[t]{.6 \textwidth}
\centering
\small
\captionof{table}{Distribution of participants across domains \\ and recommender mechanisms}
\label{tab:participant_distribution}
\begin{tabular}{@{}lcccc@{}}
\cmidrule[\heavyrulewidth]{1-4}
\multicolumn{1}{c}{\multirow{2}{*}{\textbf{Recommendation mode}}} & \multicolumn{3}{c}{\textbf{Samples per domain}} & \multicolumn{1}{c}{\multirow{2}{*}{\textbf{Total samples}}} \\
& D1 & D2 & D3 & \\ \cmidrule(lr){1-1} \cmidrule(lr){2-4} 
Automatic & 10 & 10 & 10 & 30 \\
At-End & 10 & 10 & 10 & 30 \\
On-Request & 10 & 10 & 10 & 30 \\ \cmidrule[\heavyrulewidth]{1-4}
\textbf{Total samples} & 30 & 30 & 30 & \\
\end{tabular}
\end{minipage}%
\begin{minipage}[t]{.35\textwidth}
\small
\centering
\captionof{table}{Recommendation modes per domain in each experimental group}
\label{tab:task_order}
\begin{tabular}{@{}llll@{}}
\toprule
\multicolumn{1}{c}{\multirow{2}{*}{\textbf{Group}}} & \multicolumn{3}{c}{\textbf{Recommendation mode}} \\
& D1 & D2 & D3 \\ \cmidrule(lr){1-1} \cmidrule(lr){2-4}
Group 1 & Automatic & At-End & On-Request \\
Group 2 & At-End & On-Request & Automatic \\
Group 3 & On-Request & Automatic & At-End \\ \bottomrule
\end{tabular}
\end{minipage}

\subsection{Data analysis}

\label{section:dataAnalysis}
We define in this section the analysis techniques by which we answer the research questions. 

\subsubsection{\textbf{RQ1:  \textit{What is the impact of modeling assistance on the \textbf{time} required to complete the domain models?}}}

Our objective is to precisely measure the amount of time required for each participant to complete the modeling task. Time efficiency is a direct indicator of how well the assistance is integrated with the modelers' workflow and whether it genuinely aids in accelerating the modeling process without sacrificing the depth or quality of the models. To achieve this, we employ a logging system that tracks all user operations and decisions, along with the corresponding timestamps of when they occurred.
We provide participants with 10 minutes plus any additional time required to complete their tasks, ensuring that they were not rushed. This approach reflects real-world working conditions and helps maintain the quality of the final outputs. 

We calculate the first metric \textit{Average Time to Task Completion}. This metric allows us to compare the time taken by participants to finish the modeling task across all three modes of recommendation and the unassisted mode, thereby providing insights into the efficiency of our approach.

A supplementary metric, the \textit{Number of Experiments Completed Before the Allocated Time},   evaluates the number of tasks participants successfully complete before reaching the designated time limit (10:00), serving as an indicator of the approach's efficiency in enabling participants to work without sacrificing quality.

To determine the statistical significance of the time saved between the suggestion modes and the non-assisted alternative, we perform the non-parametric Kruskal-Wallis test ~\cite{conover1999practical}. However, we do not apply this test to the "at-the-end" mode because the assistant's suggestions follow the task execution without direct assistance, resulting in no time savings.  

\subsubsection{\textbf{RQ2: \textit{What is the level of contribution of the suggestions to the resulting models?}}}

We analyze the ability of the employed approach to support modelers by providing them with pertinent elements that are approved and integrated to finalize a model under construction. 
To achieve this, we scan the obtained models of each task, and apply model transformations to derive conclusive facts, making the data more manageable. The following listing shows an example of a fact that represents an entity named \textit{Staff} and that has an attribute \textit{Name: String} and an inheritance association with the entity named \textit{Doctor}:

\begin{lstlisting}[numbers=left,caption=Example of a Derived Fact,captionpos=b,label=lst:example-CD,]
Class {name: Staff, generalizes: Doctor} 
Attributes {name: Name, type: String, class: Staff}
\end{lstlisting}

During the experiments, the logging system, integrated within MAGDA as detailed in \figref{sec:tool-logging}, tracks both the elements suggested by the recommendation system that were accepted by the user and those that were not. 
We were able then to accurately calculate  two key  metrics:
\textit{Acceptance Rate } and \textit{Contribution Rate}. Both metrics provide valuable insights into the utility and success of implementing suggestions, though from distinct perspectives. \\
The \textit{Acceptance Rate} measures the extent to which the suggested concepts are accepted by the users. The formula used for calculating it is as follows:
    
\begin{equation}
    \text{Acceptance Rate} = \frac{\text{Accepted concepts from suggestions}}{\text{Total number of suggested concepts}}\label{accuracy}
\end{equation}
    
A higher acceptance rate indicates that a larger proportion of the suggested concepts are valuable and are incorporated into the final design, whereas a lower acceptance rate means that fewer suggestions are accepted.  \\ 
The \textit{Contribution  Rate} metric, on the other hand,  provides a more general evaluation of the impact and effectiveness of the suggestions on the model resulting from the modeling task. It measures the proportion of accepted concepts from the suggestions compared to the total number of concepts in the model. To calculate it, we use the following formula (\ref{Assessment}):

\begin{equation}
    \text{Contribution Rate} = \frac{\text{Accepted concepts from suggestions}}{\text{Total number of concepts in model}}\label{Assessment}
\end{equation}

This metric provides insights into the effectiveness of the suggested concepts in improving the overall quality of the model. A higher contribution rate indicates that a larger part of the resulting model is attributable to the modeling assistant. A lower contribution rate, on the other hand, indicates that the contribution of the modeling assistant is marginal. 

Like for RQ1, we use the non-parametric Kruskal-Wallis test to assess the statistical significance of the variations in contribution levels across the various suggestion modes.

\subsubsection{ \textbf{ RQ3: Does the assistance reduce the solution \textit{diversity} among the modelers?} }

Our objective is to assess how the recommendation tool affects the creativity of participants. We do this by comparing the sequences of elements added to the model under construction by participants (concepts and attributes).  We assess the similarity of these sequences to determine the participants' level of creativity and whether the tool is restricting or enhancing their diverse and creative output.

To gain a better understanding of the diversity of solutions, we separately compare the solutions obtained for each domain-suggestion mode configuration (see \tabref{tab:task_order}). For each configuration (see Table~\ref{tab:participant_distribution}), 10 models are produced (5 in Montreal and 5 in Malaga). Thus, we have 45 pairs to compare ($C_{2}^{10}$). The overlap coefficient for a configuration is the average of the coefficients of the model pairs in that configuration.

As shown in formula ~\ref{form:overlap}, for two models containing respectively \texttt{M$_i$} and \texttt{M$_i'$} elements, the \textit{overlap} coefficient is measured as the ratio between the number of common elements and the size of the largest model. 

\begin{equation}
    \text{Overlap Coefficient} = \frac{|M_i \cap M_i'|}{\min(|M_i|, |M_i'|)}\label{form:overlap}
\end{equation}

When the calculated value is closer to 1, it indicates that the models being compared are highly similar and share a significant number of elements. Conversely, a value approaching 0 suggests that the models are vastly different, with few shared elements. This implies that participants have introduced unique concepts into each model, resulting in greater diversity.
Calculating the overlap coefficient using an exact match of elements tends to underestimate similarity, as participants may express similar concepts using different terminology. To address this limitation, we adopt a precise approach to account for this variability. Specifically, within each modeling task (Hotel, Banking, and Shopping systems), we group \texttt{semantically} equivalent elements into cohesive units, referred to as semantic "bags." For example, in our analysis, we treat "supplier" and "seller" as synonyms and place them in the same bag, illustrating this concept. When comparing two models, we consider elements within the same bag as equal. We will refer to this calculation method as the “manual” version, in contrast to the “exact” version, when presenting the results\footnote{Here, "manual" refers to manually grouping semantically equivalent elements, differing from the typical "manual vs. automated" or "semantic vs. exact" distinctions.}.

\subsubsection{\textbf{ RQ4: What kind of assistance do participants \textit{prefer to choose} when completing their tasks?}}

The final task in our study is an open-ended activity where participants can choose and switch between their preferred suggestion modes while selecting their own domain to model. This setup allows us to observe participant autonomy and gather insights into their preferences and diverse usage patterns. We specifically analyze how frequently each suggestion mode is used and how participants transition between these modes during their modeling sessions.

\subsubsection{\textbf{RQ5: {How does the assistance affect the \textit{modeling experience}, as perceived by the modelers?} }}

To address this research question, we evaluate participants' overall modeling experience using questionnaires completed after each task. These questionnaires assess various dimensions of the experience, such as ease of use, tool effectiveness, satisfaction with the process, and comfort with different recommendation modes. We analyze the quantitative responses using bar plots, which visualize the frequency and distribution of responses for each question, facilitating easy comparison across different modes. Additionally, we incorporate qualitative insights from participants' open-ended responses to provide a deeper understanding of their experiences.


\subsection{Threats to validity}\label{sec:threats}

Despite our best efforts in designing this study, several factors may limit the validity of our results.
A first threat to validity is related to the configuration of the LLM. The parameters used influence the suggestions generated. To mitigate this threat, we employed well-established settings, e.g., the \emph{text-davinci-002} engine recommended by \citet{zhao2023survey}, a balanced temperature hyperparameter~\cite{mediumGPT3Temperature}, and a reasonable setting for response length~\cite{openai2024what}.

The selection of metrics is another crucial factor. To avoid mono-method bias \cite{runeson2009guidelines}, we used two distinct metrics for questions 2 and 3. For question 3, to address the limitations of exact match methods, we opted for a manual matching approach. This method was chosen over automated tools like WordNet \cite{hearst1998automated} to prioritize context-specific categorization.

Another concern relates to the definition and execution of the modeling tasks. To ensure fairness among the suggestion modes and to counteract maturation bias, we assigned the first three tasks to participants in varying orders, as shown in \tabref{tab:task_order}. 
For the fourth task, participants were allowed to freely choose the domain and suggestion modes without specific instructions. This approach aimed to simulate a realistic scenario in which participants had the autonomy to decide what and how to model. By this stage, participants were already familiar with all three suggestion modes from the previous tasks. Additionally, this allows us to interpret the choice of suggestion mode as a genuine preference rather than mere curiosity to explore the options.

A notable threat to the validity of our study is the limited number of participants, which could affect the generalizability of our findings. However, our choice of 30 participants aligns with established guidelines in usability testing. These guidelines recommend a minimum sample size of 30 for cumulative assessments to ensure statistical significance \cite{lewis2014usability}. This number provides sufficient data for robust comparative analysis across different suggestion modes and strikes a practical balance regarding statistical power, resource allocation, time, and participant availability. To further enhance the representativeness of our results, we conducted the study with diverse groups in Montreal and Malaga.

Lastly, in the questionnaire, we did not use Likert scales with explicit point labeling. This might lead participants to interpret each level differently. We made this choice to reduce the cognitive load associated with reading and interpreting labels and to allow more flexible and nuanced interpretation between the extremes. We specifically discussed this potential threat during our pilot study and refined the questionnaire based on that discussion. Additionally, we included open-ended questions alongside the Likert scales, allowing participants to explain their choices in more detail.

\section{Results}\label{sec:results}

In this section, we present the results of the user study.

\subsection{RQ1: Effects of content assistance on time to complete task}

Table~\ref{tab:times} shows the average times required to complete the tasks using different modes for producing domain models. The standard deviations for these times are relatively low, indicating minimal variability in the data.
When comparing the \texttt{On-Request} and \texttt{Automatic} modes to the \texttt{no-assistance}, both modes demonstrate shorter average times, with reductions of 18\% and 22\%, respectively. This suggests that participants completed tasks more quickly when they could request assistance or receive automatic suggestions.
Of all the modes, the \texttt{Automatic} mode has the fastest average completion time, slightly surpassing the \texttt{On-Request} mode. This indicates that participants were most efficient when the system provided suggestions automatically as they added elements to the canvas. The slight difference between the \texttt{Automatic} and \texttt{On-Request} modes might be due to the time taken to request suggestions. 

To evaluate the statistical significance of these time differences, we performed the Kruskal-Wallis test, as detailed in Table \ref{Time:statisticaltest}. The results show that none of the p-values fall below the typical significance threshold of 0.05. This means that, although we observed time reductions with assistance modes, we cannot generalize these findings beyond our sample. The differences in average times across the modes should be interpreted cautiously, and further studies are needed to confirm these results.

\begin{table}[h!]
\caption{Comparison of task completion times across different assistance modes }
    \begin{subtable}[t]{0.6\textwidth}
        \centering
        \small
        \caption{Completion times and ratio of completed tasks}
        \label{tab:times}
        \begin{tabular}{@{}lcc@{}}
        \toprule
         & \textbf{\begin{tabular}[c]{@{}c@{}}Mean time to complete\\ the task -- min:sec (STD sec) \end{tabular}} & \textbf{\begin{tabular}[c]{@{}c@{}}Ratio of completed tasks\\ within the allocated time\end{tabular}} \\
        \midrule
        No assistance & 10:23 (20) & 0.50\\
        On-Request    & 08:29 (34) & 0.73 \\
        Automatic     & 08:07 (45) & 0.77 \\
        At-End        & 13:26 (22) & 0.03 \\
        \bottomrule
        \end{tabular}
    \end{subtable}%
    \quad
    \begin{subtable}[t]{0.3\textwidth}
        \centering
        \small
        \caption{Kruskal-Wallis statistical test Results\\}
        \label{Time:statisticaltest}

        
        \begin{tabular}{@{}lcc@{}}
        \toprule
        \textbf{Significant} & \makecell{\textbf{p-value} \\ \textbf{Mean Time} \\} \\ \midrule
        On-Request vs Automatic & \cellcolor{red!25} $0.284$ \\ 
        On-Request vs No assistance    & \cellcolor{red!25} $0.072$ \\
        Automatic vs No assistance     & \cellcolor{red!25} $0.090$ \\
        \bottomrule
        \end{tabular}
    \end{subtable}
\end{table}

In Table \ref{tab:times}, we also examine the \textit{ratio of tasks completed before the allocated time}. When assistance was provided during the modeling activity, three-quarters of the participants completed the task ahead of the allocated time. In contrast, only half of the participants managed to finish the task within the time limit without any assistance. The \texttt{At-End} mode, by its design, inherently demands more time, which explains its longer average completion time and lower completion ratio.

\begin{conclusionframe}{Answer to RQ1}
Within our sample, providing users with real-time suggestions and support during their modeling activities results in faster task completion, with a time reduction of approximately 20\% and a higher rate of tasks finished ahead of the allocated time. However, the lack of statistical significance in these results prevents us from generalizing beyond our experiment. 
\end{conclusionframe}

\subsection{RQ2: Suggestions Impact on Solution Elaboration} 
To address this research question, we assess the effectiveness of assistance provided during domain modeling tasks from two perspectives: acceptance, which measures how many suggestions were incorporated into the modeling solution, and contribution, which evaluates the portion of the resulting models attributable to these suggestions. We analyze the outcomes across different modes of assistance to understand how various types of guidance impact these metrics. The results, including their respective standard deviations, are detailed in Table~\ref{suggestionImpact-results} and discussed in Section \ref{section:dataAnalysis}.

\begin{table}[htbp]
\centering

\caption{Impact of Suggestion Modes on Solution Elaboration }

\begin{subtable}[t]{0.4\textwidth}
    \centering
    \caption{Impact of suggestions on solutions}
    \label{suggestionImpact-results}
        \small
    \begin{tabular}{@{}lcc@{}}
    \toprule
    \textbf{Mode}          & \makecell{\textbf{Acceptance} \\ \textbf{Rate (std)}} & \makecell{\textbf{Contribution} \\ \textbf{Rate (std)}} \\
    \midrule
    On-request    & 0.21  (0.12)          & 0.33     (0.11)         \\
    Automatic     & 0.33 (0.17)           & 0.56    (0.21)          \\
    At-End        & 0.11  (0.07)          & 0.08 (0.01)        \\
    \bottomrule
    \end{tabular}
\end{subtable}
\quad
\begin{subtable}[t]{0.5\textwidth}
    \centering
    
    \caption{Kruskal-Wallis statistical test Results (both metrics)  }
     \label{contribution:significanceContribution}
    \small
    \begin{tabular}{@{}lcc@{}}
    \toprule
    \textbf{Significant} & \makecell{\textbf{p-value} \\ \textbf{Acceptance Rate}} & \makecell{\textbf{p-value}  \\ \textbf{Contribution Rate}}  \\ \midrule
    On-Request vs Automatic & \cellcolor{red!25} 0.120 & \cellcolor{red!25} 0.720 \\ 
    On-Request vs At-End & \cellcolor{green!25} 0.045 &   \cellcolor{green!25} 0.022 \\
    Automatic vs At-End & \cellcolor{green!25} 0.032 & \cellcolor{green!25} 0.027  \\
    \bottomrule
    \end{tabular}
\end{subtable}

\label{tab:ContributionResults}
\captionsetup{font=footnotesize}
\end{table}

Our findings indicate that both the \texttt{On-Request} and \texttt{Automatic} modes of assistance yield higher rates of acceptance and contribution compared to providing suggestions only at the conclusion of the modeling tasks. These differences are statistically significant (see Table~\ref{tab:ContributionResults}). This suggests that delivering suggestions during the modeling process, in alignment with the current state of the domain representation, better aligns with the modelers' intentions. Conversely, offering suggestions only at the end of the task appears to be less effective. The low standard deviations of 0.07 for acceptance and 0.01 for contributions indicate that these observations are consistent across the majority of participants.

When comparing the two dynamic modes, \texttt{Automatic} assistance outperforms \texttt{On-Request} assistance in both metrics. Specifically, a third of the suggestions in the \texttt{Automatic} mode were accepted, compared to a fifth in the \texttt{On-Request} mode. Additionally, more than half of the models derived from suggestions in the \texttt{Automatic} mode, compared to a third in the \texttt{On-Request} mode. However, these differences should be interpreted with caution due to the relatively high standard deviations. 
Regarding the statistical results, the p-values show no significant difference between the \texttt{On-Request} and \texttt{Automatic} modes, despite higher averages for the \texttt{Automatic} mode. In contrast, both \texttt{On-Request vs At-End} and \texttt{Automatic vs At-End} demonstrate statistically significant differences, confirming that both \texttt{On-Request} and \texttt{Automatic} modes outperform the \texttt{At-End} mode.

\begin{conclusionframe}{Answer to RQ2}

Our user study demonstrates that suggestions provided during the modeling task are more likely to be accepted and significantly enhance the resulting models compared to those given at the end of the task. Among the dynamic assistance modes, the \texttt{Automatic} mode leads to higher acceptance and contribution rates than the \texttt{On-Request} mode. However, while these findings highlight the potential benefits of integrating real-time assistance, caution should be exercised in generalizing the differences between these two modes beyond the context of our experiment.

\end{conclusionframe}

\subsection{RQ3: Diversity of obtained models}

Based on our analysis of the domain modeling tasks across different assistance modes, we explored the similarity of the resulting models in the three domains.
We used two metrics, detailed in Section \ref{section:dataAnalysis}, to evaluate these similarities: \emph{exact} match and \emph{manual match}, with their respective standard deviations provided in Table~\ref{tab:combined-model-similarity}.

\begin{table}[htbp]

\captionsetup{font=footnotesize}
\centering

\caption{Comparison of Model Similarity for Different Tasks/Projects and Modes}

\small

\begin{tabular}{@{}llrr@{}}
\toprule
\textbf{Task/Project} & \textbf{Mode} & \makecell{\textbf{Overlap Coeff} \\ \textbf{Exact Match (std)}} & \makecell{\textbf{Overlap Coeff} \\ \textbf{Manual Match (std)}} \\
\midrule
Banking & No assistance & 0.47 (0.12)& 0.65 (0.12) \\
& On-Request & 0.48 (0.22)& 0.50 (0.31) \\
& Automatic & 0.51 (0.20)& 0.66 (0.17)\\
& At-End & 0.50 (0.21)& 0.67 (0.21)\\
\midrule
Hotel & No assistance & 0.25 (0.20) & 0.38 (0.14) \\
& On-Request & 0.46 (0.10) & 0.55 (0.12) \\
& Automatic & 0.39 (0.11) & 0.49 (0.13)\\
& At-End & 0.36 (0.13)& 0.37 (0.16) \\
\midrule
Shopping & No assistance & 0.28 (0.22) & 0.44 (0.22)\\
& On-Request & 0.37 (0.10) & 0.48 (0.10) \\
& Automatic & 0.48 (0.07) & 0.48 (0.08) \\
& At-End & 0.46 (0.17) & 0.48 (0.20) \\
\bottomrule

\end{tabular}

\label{tab:combined-model-similarity}
\end{table}
 
In evaluating the diversity of the resulting models, we consider it a proxy for creativity. A common hypothesis is that good suggestions from an assistant could lead participants to converge toward similar solutions, potentially limiting their creativity in capturing various dimensions of a domain~\cite{leach2022ai}. As shown in Table~\ref{tab:combined-model-similarity}, our findings support this hypothesis partially, as assistance tends to slightly increase the similarity of solutions when using the exact match criterion.

For the Banking domain, the similarities between the models produced with and without assistance did not show significant differences. This suggests that participants, who are presumably more familiar with the Banking domain, tend to create similar models regardless of the assistance mode used. This could be due to the structured and well-understood nature of the banking domain, which might naturally lead to a higher degree of uniformity in the models created.

In contrast, the Hotel and Shopping domains, which are broader and less familiar to participants, exhibited more pronounced similarities in the resulting models when assistance was provided. This suggests that the guidance offered by the assistance modes influenced participants to converge more closely on similar solutions in these less well-defined domains. This convergence was especially noticeable in the \texttt{Automatic} and \texttt{At-End} modes, where participants likely relied on the suggestions to navigate the varied aspects of these domains.

When considering exact matches, the overall effect of assistance on model similarity is modest but observable. However, the exact-match metric might not fully capture the nuances of semantic equivalency between concepts. Therefore, we also evaluated similarities using a manual match approach, which considers semantically equivalent concepts. Under this manual comparison, the significant difference in similarity between no-assistance and the dynamic assistance modes (\texttt{On-Request} and \texttt{Automatic}) was most evident only in the Hotel domain. This indicates that in a domain with a broader range of possible interpretations, the assistance provided by these modes may have helped participants align their models more closely, but without compromising the overall richness of the domain representation.



\begin{figure}
    \centering
    \includegraphics[width=1\columnwidth]{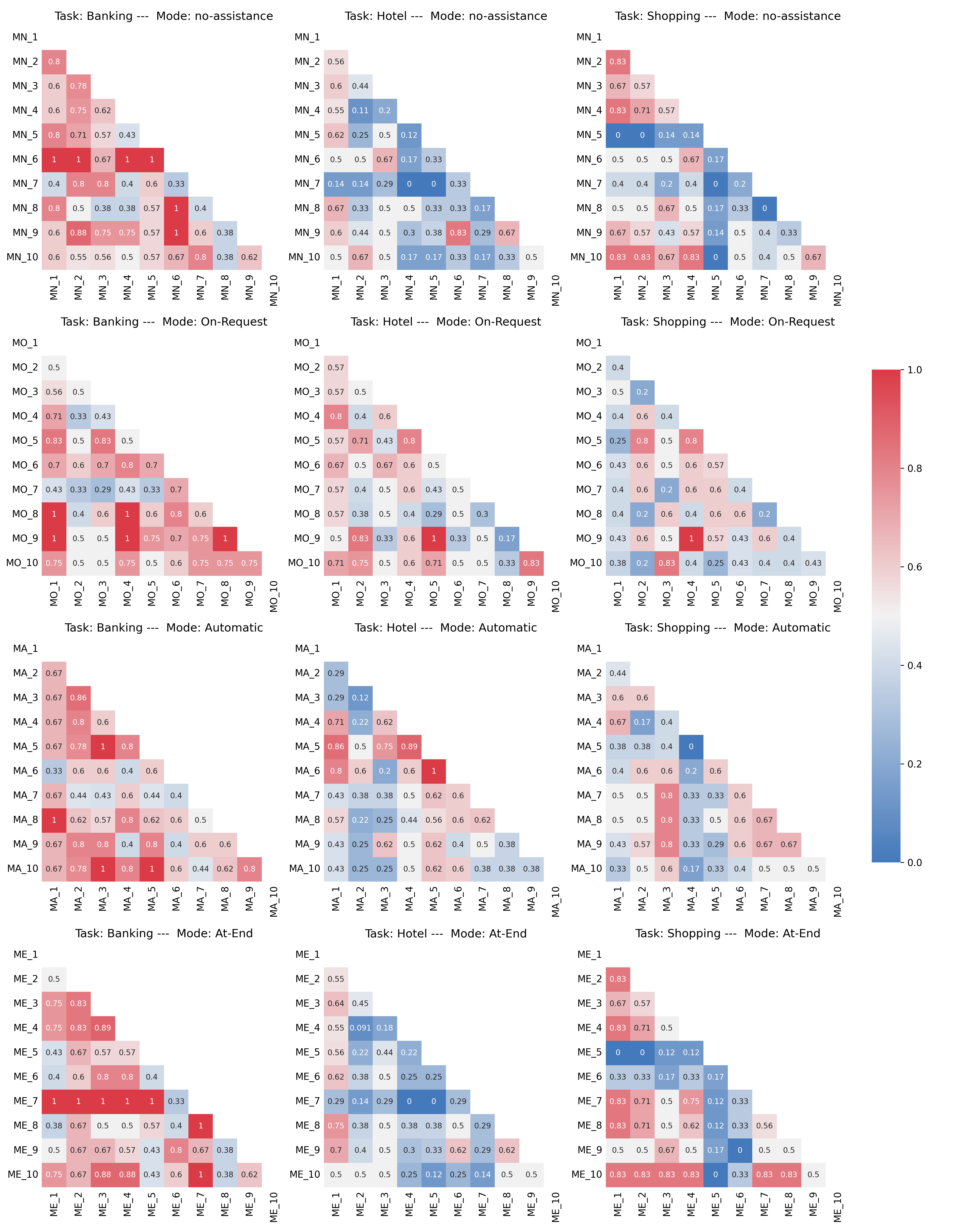}
    \caption{Similarity matrices of domain models for each task and assistance mode}
    \label{fig:MatrixSimilarity}
\end{figure}

\figref{fig:MatrixSimilarity} gives the detailed similarity matrices for pairs of models for each configuration domain-assistance mode, using the manual match.  Each matrix presents similarity scores between 0 and 1, with 1 being identical and 0 being completely dissimilar. Blue shades indicate lower similarity (closer to 0), while red shades indicate higher similarity (closer to 1).
The axes in the similarity matrices labeled as $MN\_i$, $MO\_i$, $MA\_i$, and $ME\_i$ correspond to solutions designed by participants under different assistance modes: no assistance, on-request, automatic, and at-end modes, respectively.

In the \texttt{No-Assistance} Mode, model similarities are more variable across the tasks (Hotel and Shopping systems), indicating that users explore a wider range of solutions. However, in the Banking task, some models still show high similarity, suggesting that users may naturally converge on similar solutions even without guidance.

In the \texttt{On-Request} Mode, model similarities are generally higher, particularly in the Banking and Hotel tasks, indicating that users tend to create more similar solutions when seeking specific guidance. However, in the Shopping task, the improvement over the non-assistance option is modest, suggesting that this mode is less effective in guiding users toward uniform solutions in that context.

In the \texttt{Automatic}  Mode, the highest similarities are observed across models ($MA\_i$), in Banking system, indicating that continuous, real-time guidance leads to more consistent outcomes. For Hotel and Shopping tasks, where variability is usually higher, similarities improved compared to the non-assistance option, demonstrating that the automatic suggestions steer users toward uniform solutions.

In the \texttt{At-End} Mode, similarities fall between those in the No Assistance and On Request modes. Users initially create models independently, resulting in some variation, but final suggestions lead to partial convergence. This pattern, seen across all tasks, shows that end-of-process feedback helps refine and align models, though not as strongly as real-time suggestions.

The nature of the task influences the impact of assistance. More structured tasks like Banking system, show higher similarities, even without assistance, while less structured tasks like Shopping exhibit greater variability, especially without or with minimal assistance.

\begin{conclusionframe}{Answer to RQ3}

In our sample, assistance during modeling had varying effects on solution similarity depending on the domain and the mode of assistance. The differences compared to the non-assisted condition were not substantial enough to conclude that assistance restricts the creativity of modelers.

\end{conclusionframe}

\subsection{RQ4: Participant preference analysis}

\begin{figure}[h!]
    \centering
    
    \begin{subfigure}[t]{0.48\textwidth}
        \centering
        \includegraphics[width=\textwidth]{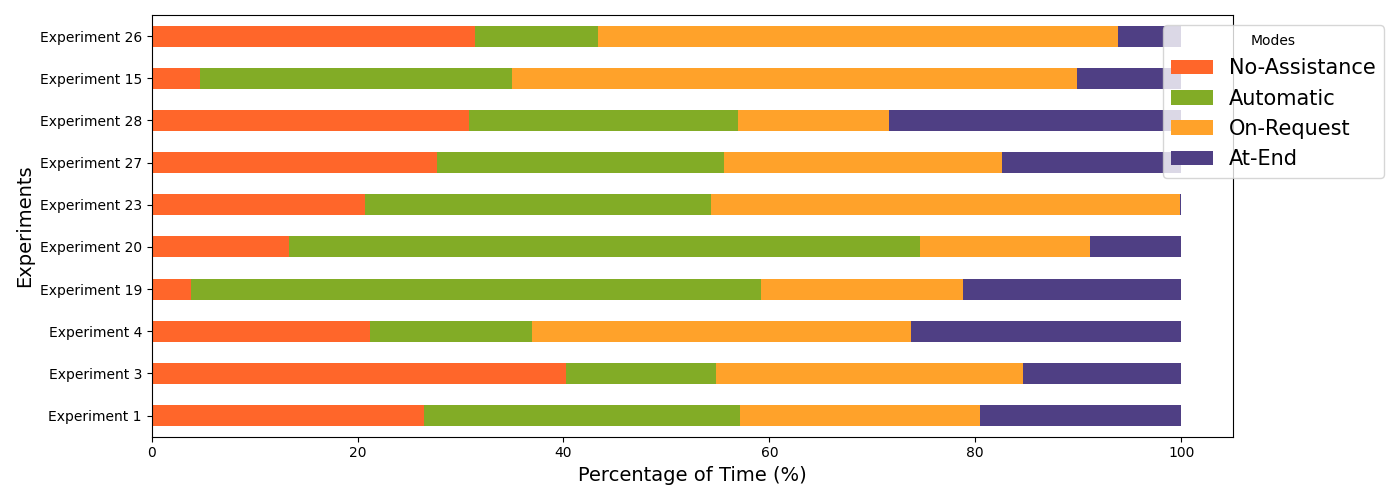}
        \caption{Mode Sequences: No-Assistance, Autom., On-Request, At-End}
        \label{fig:combined_complete_group_1}
    \end{subfigure}
    \hfill
    \begin{subfigure}[t]{0.48\textwidth}
        \centering
        \includegraphics[width=\textwidth]{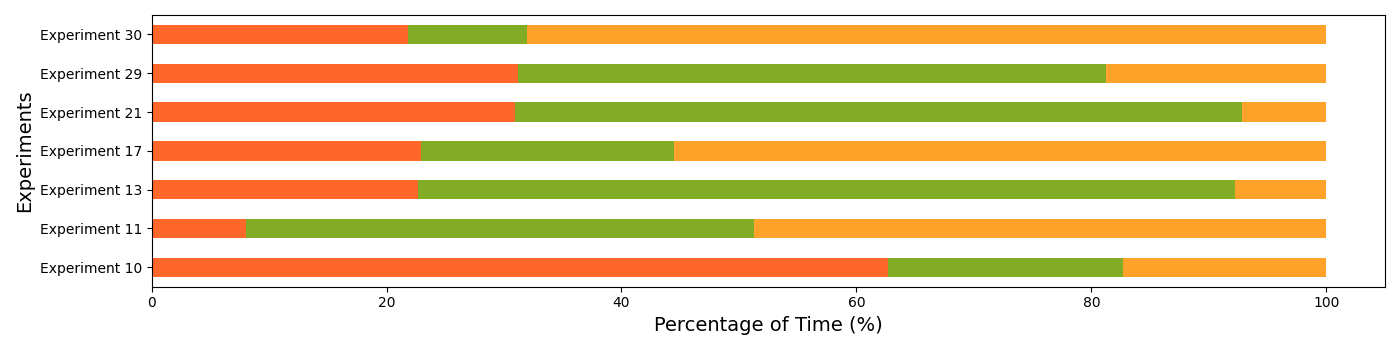}
        \caption{Mode Sequences: No-Assistance, Automatic, On-Request}
        \label{fig:combined_incomplete_group_1}
    \end{subfigure}
    
    \begin{subfigure}[t]{0.48\textwidth}
        \centering
        \includegraphics[width=\textwidth]{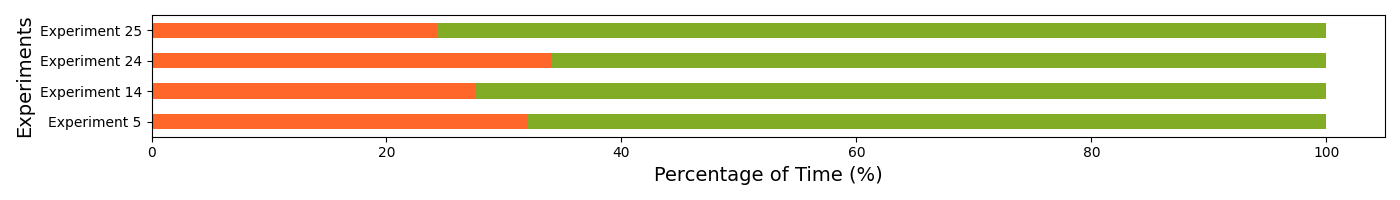}
        \caption{Mode Sequences: No-Assistance, Automatic}
        \label{fig:remaining_complete_group_2}
    \end{subfigure}
    \hfill
    \begin{subfigure}[t]{0.48\textwidth}
        \centering
        \includegraphics[width=\textwidth]{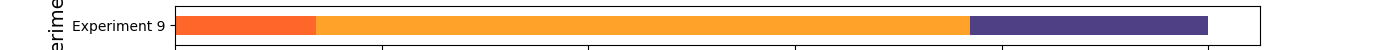}
        \caption{Mode Sequences: No-Assistance, Automatic At-End}
        \label{fig:remaining_complete_group_3}
    \end{subfigure}
    
    \begin{subfigure}[t]{0.48\textwidth}
        \centering
        \includegraphics[width=\textwidth]{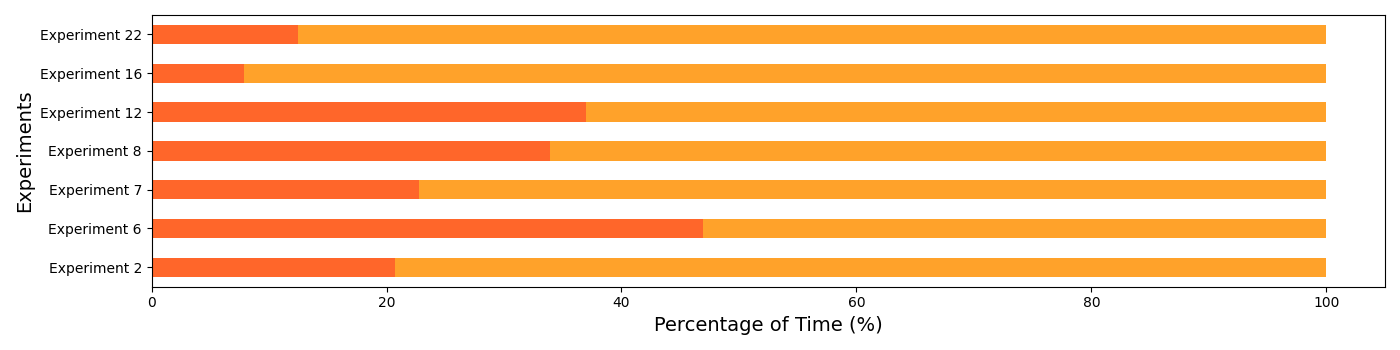}
        \caption{Mode Sequences: No-Assistance,  On-Request}
        \label{fig:remaining_incomplete_group_1}
    \end{subfigure}
    \hfill
    \begin{subfigure}[t]{0.48\textwidth}
        \centering
        \includegraphics[width=\textwidth]{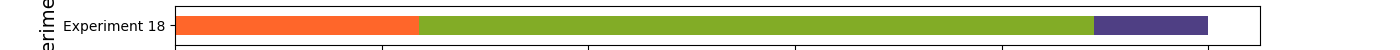}
        \caption{Mode Sequences: No-Assistance, Automatic,  At-End}
        \label{fig:remaining_incomplete_group_2}
    \end{subfigure}
    \caption{Mode transition patterns in various assistance modes. Bars represent experiments, colors to show the proportion of time spent in 'No-assistance' (orange), 'Automatic' (green), 'On-Request' (yellow), and 'At-End' (purple) modes.}
   
\label{fig:mode-utilsation}
\end{figure}

We analyzed participant behavior while they used the tool to model a domain of their choice without specific instructions on which assistance mode to use. Our objective is to understand how the provided assistance impacts user choices, uncovering insights into modelers' preferences and their interactions with the tool's different modes when given complete autonomy.
\figref{fig:mode-utilsation} presents an overview of the time spent by each of the 30 participants in each mode and visualizes their transitions between these modes. Initially, participants spend time building a partial model (shown in red). Then, they switch to an assistance mode for a certain duration (indicated by the color corresponding to the mode). They may either remain in that mode or switch to another. The subfigures show different patterns of mode transitions.

\begin{figure}
    \centering
    \includegraphics[width=0.8\linewidth]{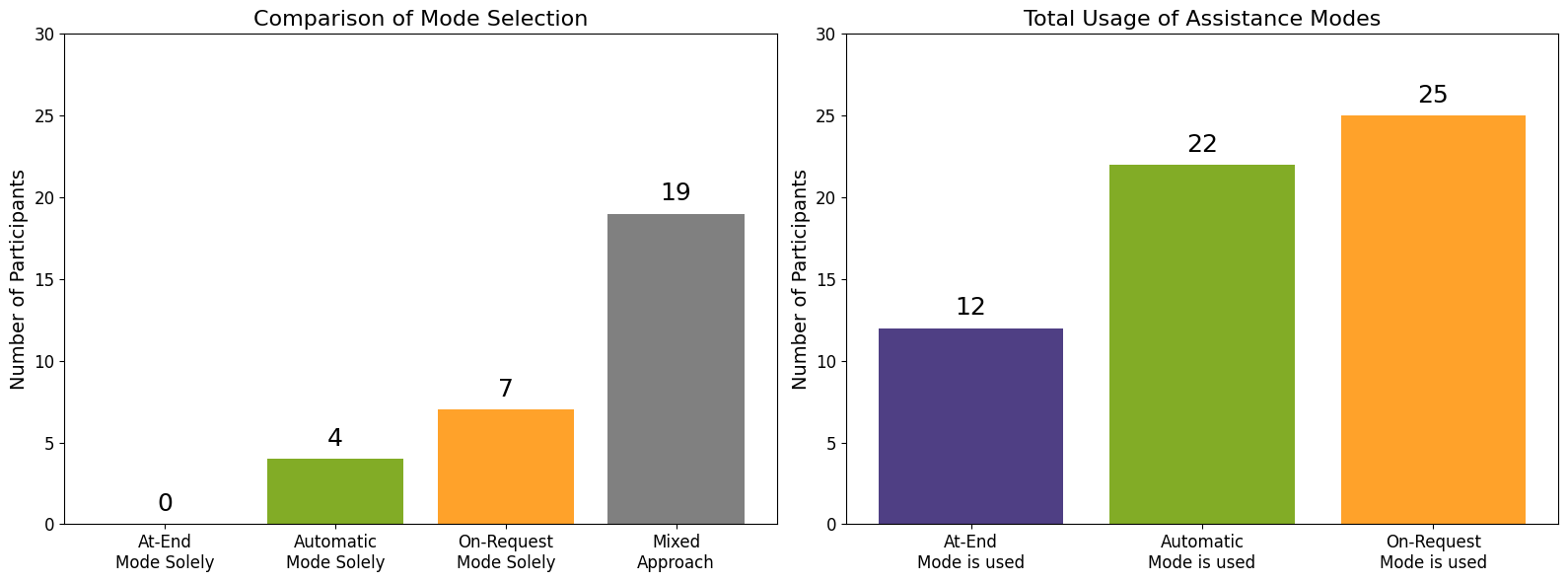}
    \caption{Participant Preferences and Usage Patterns for Assistance Modes}
    \label{fig:participants preferences}
\end{figure}

\paragraph{Diversity in Mode Selection:} \figref{fig:mode-utilsation} shows significant variability in the time it takes for participants to begin using the provided assistance after constructing a partial model. This ranges from less than 10\% to more than 60\% of the total session time. This variability can be influenced by several factors. First, individuals have varying levels of autonomy; some prefer to solve problems independently before seeking help, while others may engage with assistance more readily. Second, there is a spectrum of willingness to ask for help, which can be influenced by personal preferences. Some individuals are more comfortable or inclined to seek assistance, while others view it as a last resort. Finally, the perceived usefulness of  AI assistance is crucial. Confidence in the AI’s capability to provide meaningful and helpful support varies widely among participants, affecting how quickly they turn to it as a resource.

The analysis also reveals diverse behaviors in the selection of assistance modes, as shown in \figref{fig:participants preferences}. A few participants exclusively used a single mode—specifically, 4 for \texttt{Automatic} and 7 for \texttt{On-Request}. The majority of participants (19) adopted a \textit{mixed} approach, using different modes at various points during the modeling session. We hypothesize that the need for assistance, and the type of assistance required, depends on the state of the modeling process. Additionally, participants significantly preferred modes that provide suggestions throughout the modeling process, both in terms of frequency and time spent, compared to the \texttt{At-End} mode, which was used by only a third of the participants.

    \paragraph{Mode Transition Patterns:} 
    
Our study of mode transitions reveals interesting patterns in how participants interact with the tool during experiments. These patterns can be observed in \figref{fig:mode-utilsation} and are summarized in Table \ref{tab:mode_transitions}. The two prominent findings concern the start of the assistance and transitions between modes.

\textbf{Frequent Start with \texttt{Automatic} Mode:} The majority of participants start the assistance using the \texttt{Automatic} mode(22), with most of them transitioning to other modes later on (18) or continuing without transitioning until the session's end (4). This common choice suggests that users initially value the automatic, context-sensitive suggestions provided by this mode, finding it useful for kick-starting their engagement with the system's functionalities.

\textbf{Common Path from  
 \texttt{Automatic} Mode to \texttt{On-Request} Mode:} Despite the flexibility to switch between modes, it is noteworthy that none of the 8 participants who started with the \texttt{On-Request} mode transitioned to the \texttt{Automatic} mode. This indicates that those who choose the \texttt{On-Request} mode do so with a clear intention, likely preferring a more controlled form of assistance. Conversely, participants frequently transition from the \texttt{Automatic} mode to the \texttt{On-Request} mode. This behavior can be explained by the decreasing relevance of automatic suggestions as the modeling process evolves. Additionally, as the domain becomes better represented, modelers may seek more targeted assistance to explore specific aspects rather than receiving general suggestions, thus favoring the \texttt{On-Request} mode.

\begin{table}[h!]
  \centering
  
  \caption{Transitions between assistance modes. Rows indicate switches between modes and the frequency of these transitions.}
  \small
  \begin{tabular}{llr}
    \toprule
    From Mode & To Mode & Count \\
    \midrule
    No-Assistance & Automatic & 22 \\
    Automatic & OnRequest & 17 \\
    OnRequest & At-End & 10 \\
    No-Assistance & OnRequest & 8 \\

    Automatic & At-End & 1 \\
    OnRequest & Automatic & 0 \\

    \bottomrule
  \end{tabular}
\label{tab:mode_transitions}
  
\end{table}

\begin{figure}
\begin{conclusionframe}{Answer to RQ4}
When users are given complete autonomy, they typically prefer to begin by receiving automatic suggestions, which helps them get started with the modeling process. As they progress and their domain representations become more developed, they tend to seek more targeted suggestions to refine and complete their models. There is generally less interest in receiving additional suggestions after the modeling task is completed.
\end{conclusionframe}
\end{figure}

\subsection{RQ5: Impact of Modeling Assistance on User Experience}

In this section, we present the results of the questionnaire that each participant completed after finishing the tasks. The questionnaire consisted of two parts: closed-ended questions and open-ended questions. For the first part, we used a 1 to 5 agreement scale, with 1 being the lowest value and 5 the highest, which is commonly employed in research involving questionnaires. This scale allowed us to evaluate the participants' responses effectively.
This part of the questionnaire consisted of various statements aimed at capturing participants' experiences and perceptions during the experiment. Notably, all the questions have positive statements, with one or two negative statements included as suggested by \citet{saris2014design}. This is to ensure that participants were reading carefully and providing thoughtful responses rather than selecting answers randomly. For example, in the question on tool characteristics, \textit{Easy to use}, \textit{Good fit for the purpose}, \textit{Intuitive}, and \textit{Visually appealing} were positive statements, while \textit{Slow} served as the negative one.

\begin{figure}[h]
    \centering
    \captionsetup{font=footnotesize}
    \begin{subfigure}{0.32\linewidth}
        \centering
        \includegraphics[width=\textwidth]{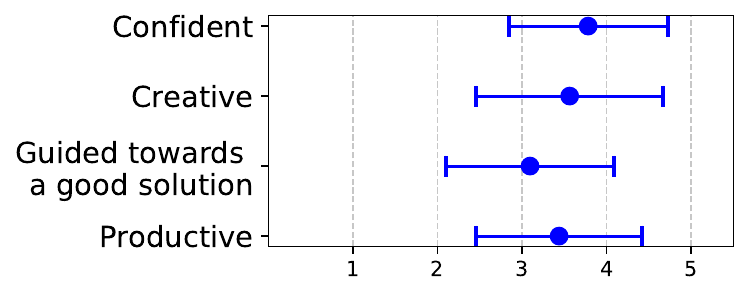}
        \caption{On-Request Mode}
        \label{fig:imageOnRequest-duringExp}
         \Description{Invalid}
    \end{subfigure}
    \hfill     
    \begin{subfigure}{0.32\linewidth}
        \centering
        \includegraphics[width=\textwidth]{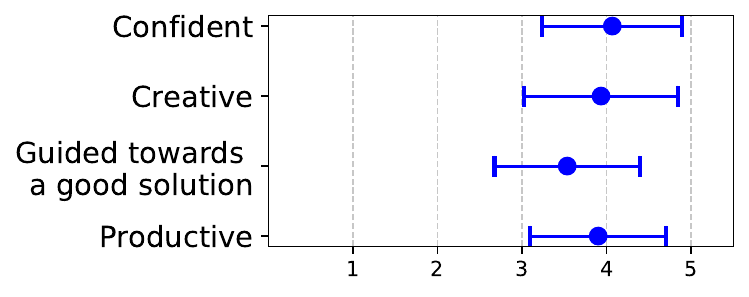}
        \caption{Automatic Mode}
          \Description{Invalid}
        \label{fig:image3-automatic}
    \end{subfigure}
       \hfill
    \begin{subfigure}{0.32\linewidth}
        \centering
        \includegraphics[width=\textwidth]{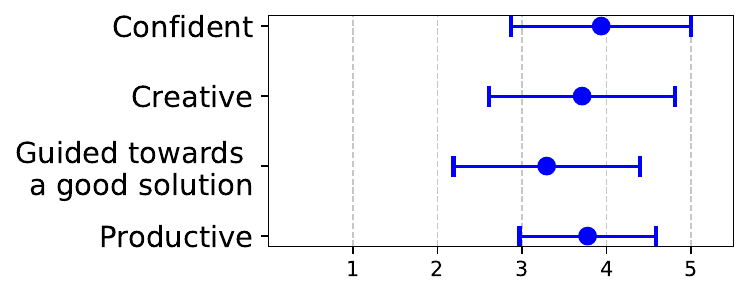}
        \caption{At-End Mode}
          \Description{Invalid}
        \label{fig:image2-atEnd}
    \end{subfigure}
        \begin{subfigure}{0.32\linewidth}
        \centering
        \includegraphics[width=\textwidth]{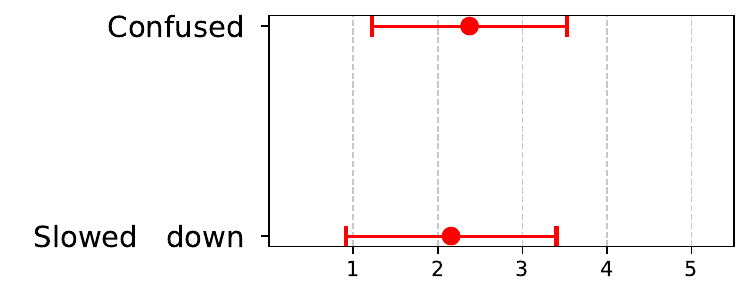} 
        \caption{On-Request mode  (negative adjectives)}
          \Description{Invalid}
        \label{fig:image4-onRequest}
    \end{subfigure}
    \hfill
    \begin{subfigure}{0.32\linewidth}
        \centering
        \includegraphics[width=\textwidth]{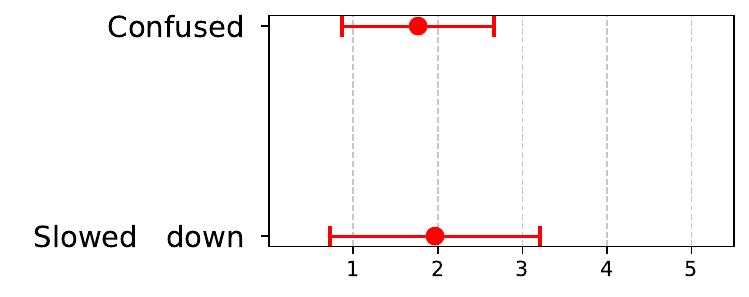} 
        \caption{Automatic mode  (negative adjectives)}
          \Description{Invalid}
        \label{fig:image6-aut}
    \end{subfigure}
      \hfill
    \begin{subfigure}{0.32\linewidth}
        \centering
        \includegraphics[width=\textwidth]{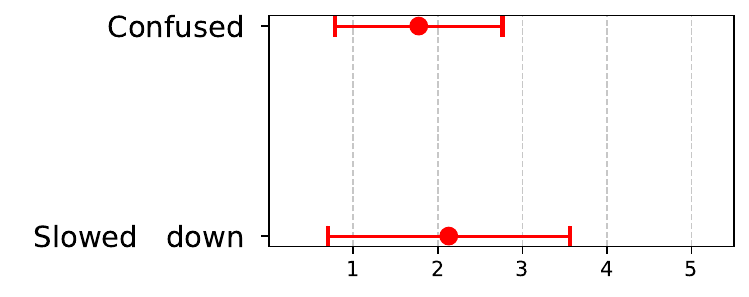} 
        \caption{At-End mode (negative adjectives)}
          \Description{Invalid}
        \label{fig:image5-atEnd-duringexp}
    \end{subfigure}
    \caption{Answers to  question: \textit{During the experiment, I felt...}}
    \label{fig:duringExperiment}
\end{figure}

\smallsection{Results regarding participants' feeling }
The analysis of participant responses, as shown in \figref{fig:duringExperiment}, across different assistance modes reveals generally positive sentiments, with mean scores of 3 or higher across all modes. \textit{Confidence} levels are consistent across the modes, with mean values close to 4, indicating participants felt similarly confident throughout. The \textit{creative} statement received higher ratings in the \texttt{Automatic} mode, with a smaller standard deviation (a smaller bar), indicating less variability and a generally more positive perception. A similar trend is observed for the \textit{Productive} statement.
Likewise, the \textit{Guided towards a good solution} statement scored notably higher in the \texttt{Automatic} mode, with an average close to 3.5 and a low standard deviation compared to other modes. Overall, while all modes were positively received, participants showed a preference for the \texttt{Automatic} mode.

For the negative statements, the consistently low scores in \textit{confusion} and \textit{slowed down} across all modes highlight the user-friendly nature of the assistance provided. Despite slight variations between the modes, the assistance does not appear to negatively affect the modeling task experience.

\begin{figure}[h]
    \centering
    \captionsetup{font=footnotesize}
    \begin{subfigure}{0.32\linewidth}
        \centering
        \includegraphics[width=\textwidth]{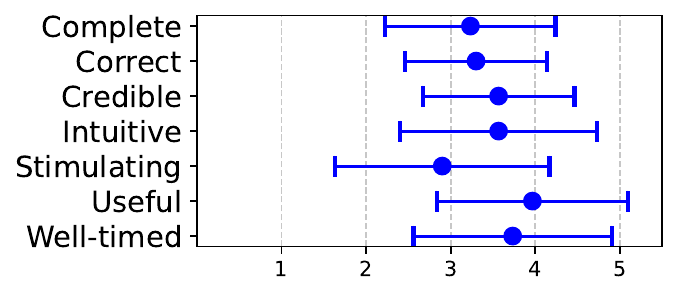}
        \caption{On-Request Mode}
          \Description{Invalid}
        \label{fig:imageOnRequest-recom}
    \end{subfigure}
    \hfill
    \begin{subfigure}{0.32\linewidth}
        \centering
        \includegraphics[width=\textwidth]{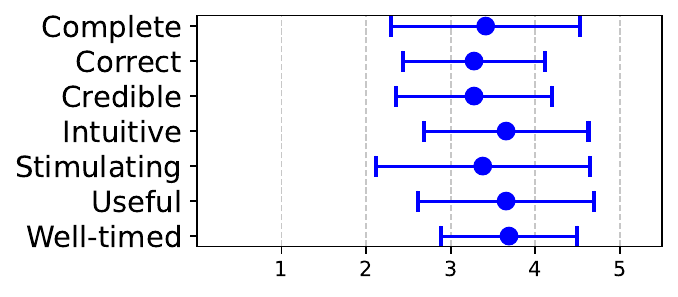}
        \caption{Automatic Mode}
          \Description{Invalid}
        \label{fig:image3-autom-recom}
    \end{subfigure}
    \hfill
    \begin{subfigure}{0.32\linewidth}
        \centering
        \includegraphics[width=\textwidth]{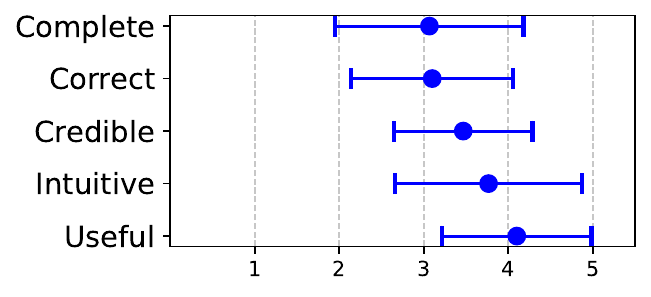}
        \caption{At-End Mode}
          \Description{Invalid}
        \label{fig:image2-atEnd-rec}
    \end{subfigure}
    \begin{subfigure}{0.32\linewidth}
        \centering
        \includegraphics[width=\textwidth]{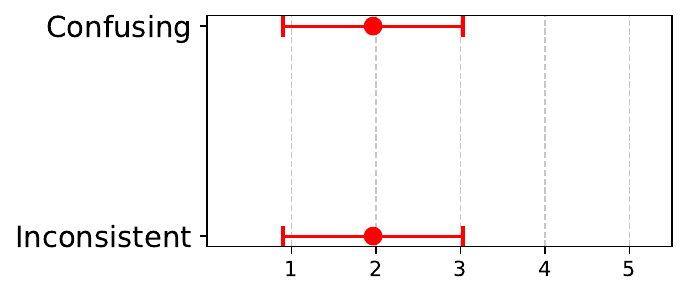} 
        \caption{On-Request mode  (negative adjectives)}
          \Description{Invalid}
        \label{fig:image4-onreq-recom-negatif}
    \end{subfigure}
    \hfill
    \begin{subfigure}{0.32\linewidth}
        \centering
        \includegraphics[width=\textwidth]{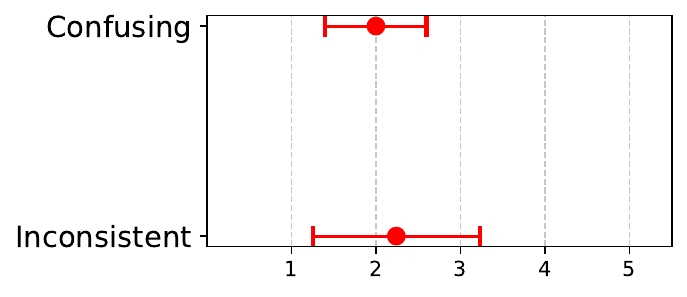} 
        \caption{Automatic mode  (negative adjectives)}
          \Description{Invalid}
        \label{fig:image6-automatic-recomm-negatif}
    \end{subfigure}
    \hfill
    \begin{subfigure}{0.32\linewidth}
        \centering
        \includegraphics[width=\textwidth]{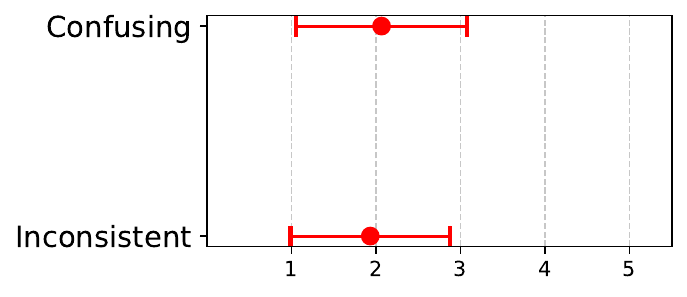} 
        \caption{At-End mode (negative adjectives)}
          \Description{Invalid}
        \label{fig:image5-atEnd-recom-negatif}
    \end{subfigure}
 
    \caption{Answers to  question: \textit{I found the recommendations...}}
   \Description{Invalid}
    \label{fig:recomendations}
\end{figure}

\smallsection{Results regarding the recommendations} \figref{fig:recomendations} presents participant perceptions of recommendation effectiveness across the three modes. Overall, positive statements scored highly, with mean values between 3 and 4 for all modes. The \texttt{At-End} and \texttt{On-Request} modes were rated particularly high for their usefulness, consistent with their design to provide feedback under the modeler's control. This somewhat contradicts the objective findings of RQ2, where the acceptance and contribution rates for the \texttt{At-End} mode were found to be low. On the other hand, the \textit{completeness} and \textit{correctness} statements for the \texttt{At-End} mode received lower scores compared to the other modes, which aligns with the RQ2 results.
Additionally, we found that the recommendations provided in the \texttt{Automatic} mode were considered more stimulating than those in the \texttt{On-Request} mode. It is important to note that the \textit{Stimulating} and \textit{Well-Timed} statements were not presented to participants for the \texttt{At-End} mode, as this mode only offers recommendations after the modeling task is complete.
For the remaining statements, no significant trends were observed.

Regarding negative statements, all three modes showed low levels of \textit{confusion} and \textit{inconsistency}, with the \texttt{Automatic} mode exhibiting slightly higher inconsistency. While the mean scores for \textit{Confusing} were similar across modes, the \texttt{Automatic} mode had a notably lower standard deviation.

\begin{figure}[h]
    \centering
    \captionsetup{font=footnotesize}
    \begin{subfigure}{0.32\linewidth}
        \centering
        \includegraphics[width=\textwidth]{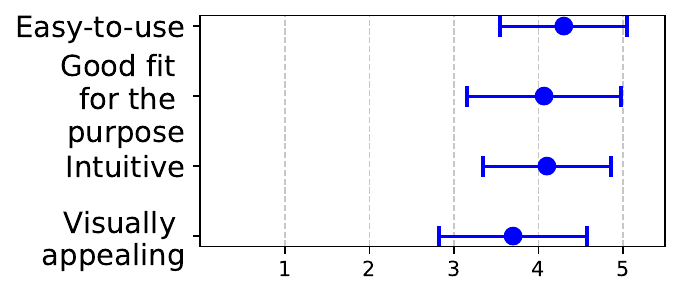}
        \caption{On-Request Mode}
          \Description{Invalid}
        \label{fig:imageOnRequest}
    \end{subfigure}
    \hfill
    \begin{subfigure}{0.32\linewidth}
        \centering
        \includegraphics[width=\textwidth]{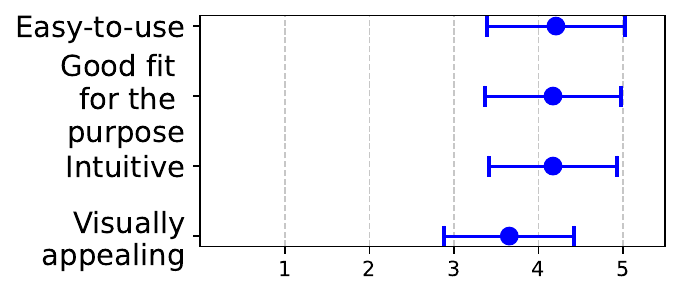}
        \caption{Automatic Mode}
          \Description{Invalid}
        \label{fig:image3-automaticTool}
    \end{subfigure}
    \hfill
    \begin{subfigure}{0.32\linewidth}
        \centering
        \includegraphics[width=\textwidth]{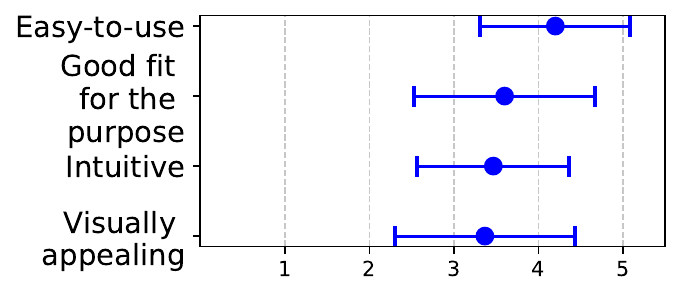 }
        \caption{At-End Mode}
          \Description{Invalid}
        \label{fig:image2-atEndTool}
    \end{subfigure}
        \begin{subfigure}{0.32\linewidth}
        \centering
        \includegraphics[width=\textwidth]{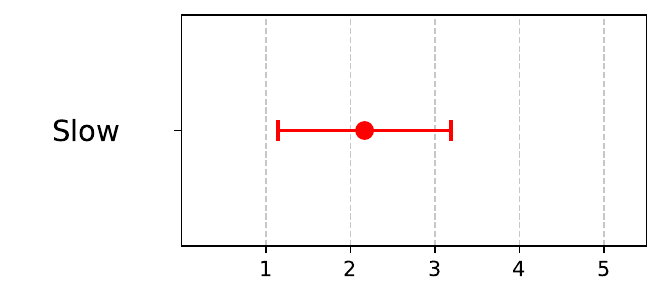} 
        \caption{On-Request mode  (negative adjectives)}
          \Description{Invalid}

        \label{fig:image4-onRequesTool}
    \end{subfigure}
    \hfill
    \begin{subfigure}{0.32\linewidth}
        \centering
        \includegraphics[width=\textwidth]{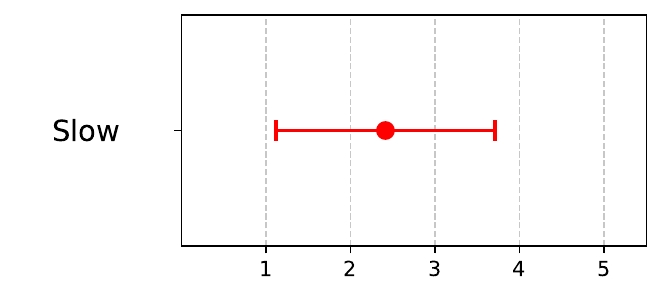} 
        \caption{Automatic mode  (negative adjectives)}
        \label{fig:image6automatic-tool-negatif}
    \end{subfigure}
        \hfill
    \begin{subfigure}{0.32\linewidth}
        \centering
        \includegraphics[width=\textwidth]{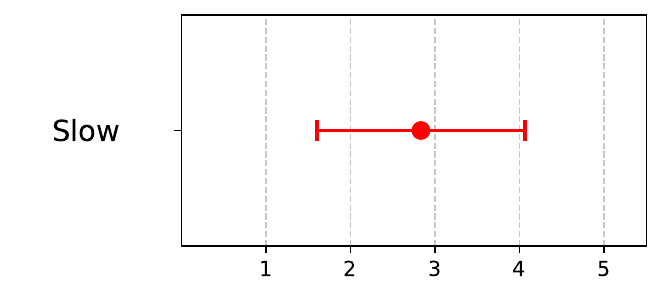} 
        \caption{At-End mode (negative adjectives)}
          \Description{Invalid}

        \label{fig:image5-atEndToolNegatif}
    \end{subfigure}
    \caption{Answers to  question: \textit{I found the tool...}}
    \label{fig:results-tool}
\end{figure}

\smallsection{Results regarding the tool} \figref{fig:results-tool} presents a comparative analysis of participants' responses regarding their experience with the tool across the three modes. For the \textit{easy-to-use} statement, participants gave high scores to all three modes, indicating that incorporating assistance does not seem to complicate the modeling environment overly. However, the way the tool presents recommendations in the \texttt{At-End} mode appears to be less of a \textit{good fit for the purpose} and less \textit{intuitive} compared to the two continuous assistance modes. This suggests that presenting a large number of recommendations at the end should be approached differently than providing recommendations progressively during the modeling task. The final positive statement, \textit{visually appealing}, received slightly lower scores, but still above 3.5 average for the continuous modes. This feedback calls for dedicating more effort to improving the aesthetics of the tool.

For the negative statement \textit{slow}, participants gave average scores close to 2 for both the \texttt{Automatic} and \texttt{On-Request} modes, with a slight preference for the latter. The \texttt{At-End} mode was perceived slower, with an average score closer to 3.

\smallsection{Feedback from open-ended questions} 
In the second part of the questionnaire, we included three open-ended questions: (1) \textit{Which features would you improve, and how?}, (2) \textit{What features did you miss?}, and (3) \textit{Do you have any other remarks?} Since participants answered these questions after each modeling task, we expected a maximum of 270 responses (30 participants × 3 tasks × 3 questions). However, as responding was optional, not all participants provided answers to every question. Despite this, we gathered a substantial amount of feedback, allowing for a comprehensive categorization of participants’ input. This feedback is crucial for understanding diverse perspectives and experiences with the modeling assistance tool. 

Responses to the first question, \textit{Which features would you improve, and how?}, highlighted specific areas for enhancement and provided insight into user satisfaction with various aspects of the tool. Table \ref{tab:user_feedback} shows the frequency of feature groups mentioned in the responses. \begin{table}[ht]
\centering
\small

\caption{Categorization of Participant Feedback about Implemented Features}

\begin{tabular}{@{}p{5cm}p{8cm}r@{}}
\toprule
\textbf{Category} & \textbf{Description} & \textbf{Count} \\ \midrule

User Interface and Usability & Enhancements to UI and usability, such as  
readability, navigation, visual elements, and interface intuitiveness. & 22 \\
Quality of association recommendations and its management & Improvements in association recommendations, including better naming, reducing redundancy, and handling aggregation/compositions/inheritance. & 20 \\

Performance and Efficiency & Feedback on the speed and responsiveness of the tool, including load times and handling multiple suggestions. & 18 \\
Suggestions of other system features & Specific features of the recommendation system needing improvement. & 16 \\

 Quality of class names and attributes Recommendations & Quality and relevance of the class names and attribute recommendations, including alignment with tasks and specificity. & 15 \\

General Feedback and Miscellaneous & General observations and miscellaneous suggestions that provide insights into user experience and potential improvements. & 9 \\ \bottomrule
\end{tabular}

\label{tab:user_feedback}

\end{table} 
Upon analyzing the feedback, we found that many participants emphasized improving the interface and usability, particularly in how recommendations are presented, especially for associations. Enhancing performance and efficiency were also frequently suggested.

Regarding the second question, \textit{What features did you miss?}, responses varied. They highlighted several areas for potential improvement.  Table \ref{tab:missing_features} shows the categories of features that were mentioned together with their frequencies.

\begin{table}[ht]
\small
\centering

\caption{Categorization of Missing Features Based on Participant Feedback}

\begin{tabular}{@{}p{3.5cm}p{9.5cm}r@{}}
\toprule
\textbf{Category} & \textbf{Description} & \textbf{Count} \\ \midrule
Attribute Enhancements & Suggestions related to improving attribute features, including adding new attribute suggestions, handling attributes in automatic modes, and better recommendations for attribute types. & 23 \\
Class  names Enhancements & Suggestions for improving class names  & 12 \\

User Interface Controls & Feedback on user interface elements such as adding decline or cancel buttons, improving the ease of rejecting suggestions, and organizing tabular data. & 11 \\
Association Improvements & Enhancements to association features like inferring cardinalities, managing bidirectional associations, and better naming for associations. & 10 \\

Miscellaneous & Diverse suggestions that did not fit into other categories but were mentioned by participants, such as adding methods. & 10 \\ 

General Usability & General suggestions for improving usability, such as on-hover descriptions, step-by-step progress visibility, and the ability to undo actions. & 9 \\

\bottomrule
\end{tabular}

\label{tab:missing_features}
\end{table}
The feedback highlights that participants primarily missed features related to the \textit{Attribute} construct in class diagrams. In the design of the assistance tool, the focus was placed on the main construct, i.e., \textit{Class}, while attributes received less attention. This brings up the challenge of providing recommendations when construct types are interdependent. In a similar vein, missing features related to classes and other types of relationships were mentioned, though less frequently. Another noteworthy category of missing features concerns interface controls. For example, some participants requested decline or cancel buttons to facilitate smoother interactions with the tool. This feedback suggests that the success of intelligent assistance depends not only on the quality of recommendations but also on how they are presented to the user.  It offers valuable insights that not only guide improvements to the class diagram assistant but also provide recommendations for developing assistants in other modeling languages.

Participants' responses to the third question, \textit{"Do you have any other remarks?"}, are summarized in Table \ref{tab:other_remarks}. Once again, the remarks are grouped into thematic categories, capturing both general sentiment and specific feedback.

\begin{table}[ht]
\centering
\small

\caption{Categorization of Additional Remarks Based on Participant Feedback}
\begin{tabular}{@{}p{3.5cm}p{9.5cm}r@{}}
\toprule
\textbf{Category} & \textbf{Description} & \textbf{Count} \\ \midrule
General Approval & Positive feedback appreciating the tool's usefulness or specific features. Includes comments on usability, helpfulness, and overall satisfaction. & 18 \\

Enhancement Suggestions & Suggestions for enhancing the tool's functionality, including adding specific features like domain specification or improving the suggestion mechanism. & 7 \\ 

User Interface and Usability & Feedback focusing on the user interface aspects such as ease of use, layout issues, and suggestions for UI improvements like better representation of suggestions. & 6 \\
Performance Concerns & Concerns regarding the tool's performance, particularly its speed and responsiveness during operation. & 5 \\

Miscellaneous & Insights into user experience or potential improvements. & 5 \\ \bottomrule
\end{tabular}

\label{tab:other_remarks}

\end{table}

The feedback from the additional remarks reflects a general approval of the tool, with most participants expressing satisfaction regarding its usability and usefulness. Some participants used this opportunity to suggest additional improvements to the tool’s functionality and user interface, while a few raised concerns about performance issues. Addressing these enhancement suggestions and performance concerns could further increase overall satisfaction and the tool’s effectiveness.

As a final analysis step, we explored the potential relationship between participants' backgrounds (position and level of education) and their responses to the open-ended questions. However, we did not identify any trends indicating that certain categories of suggestions were influenced by participants' backgrounds.

\begin{conclusionframe}{Answer to RQ5}
Assistance significantly enhances the modeling experience, with participants valuing the support provided by the LLM-based tool. Nonetheless, the feedback highlights areas for improvement, particularly in the relevance of suggestions, user experience, performance, and visual representations, to better support domain modeling tasks.

\end{conclusionframe}
    
\section{Discussion}\label{sec:discussion}
In this section, we provide a more general discussion about the results gathered in Section \ref{sec:results}, focusing on both the outcomes of the solutions generated and concerns related to user experiences. 
We explore how these observations align with the effectiveness of modeling assistance.
We also reflect on the implications of these findings for tool developers, aiming to derive actionable insights that can inform future enhancements in software modeling assistance.

\subsection{Technical Perspective}  
The introduction of a domain modeling assistance component utilizing few-shot learning to recommend missing elements for incomplete models has significantly impacted both the quality and efficiency of the solutions produced. Our results demonstrate that such assistance notably reduces the time required to complete tasks across various recommendation modes, indicating a clear enhancement in productivity.

The increased efficiency does not compromise solution quality. On the contrary, the high acceptance and contribution rates of suggestions hint at relevant recommendations that contribute positively to the models being developed. This underscores the effectiveness of the assistance component in improving the modeling process.

However, while MAGDA excels in suggesting concepts, the performance in recommending relationships is more variable. Inherence relationships are well captured, yet associations and compositions show a higher rate of inaccurate suggestions. This discrepancy highlights areas where the component's predictive capabilities could be further refined.

\subsection{Human Perspective} 
The analysis of participant behavior during the free modeling task, along with their feedback, reveals several insights. Participants showed a preference for using assistance and selected both \texttt{on-Request} and \texttt{automatic} modes. This behavior indicates a desire for a balance between autonomy and guided assistance. The \texttt{on-Request} mode offers control over when to receive help, aligning with users' need for agency in the design process. Conversely, the \texttt{automatic} mode might be preferred for its efficiency and ability to introduce unexpected yet relevant ideas. The high acceptability of suggestions indicates that users find the LLM-generated options relevant and useful, likely due to the LLMs' ability to generate contextually appropriate recommendations, thereby enhancing the user's design process.

Participant feedback after carrying out the experiments suggests areas for improvement, including performance, recommendation management, and the inclusion of additional features. These insights can guide the further development and refinement of the tool to enhance its overall effectiveness and user satisfaction.

Reflecting on the utility of our approach, it is evident that integrating assistants that use few-shot learning into software modeling tasks enhances productivity, creativity, and the user experience through tailored assistance. Participants were receptive to this assistance during the modeling task, highlighting their readiness to adopt advanced support systems.

These findings underscore the potential of AI techniques in general, and LLMs in particular, to transform traditional tasks by bridging the gap between human intuition and machine efficiency. This suggests a promising future for intelligent-enhanced design tools in software engineering.

\subsection{Tooling Perspective}
The development of MAGDA showcases the significant benefits of integrating intelligent assistance into existing modeling environments. By leveraging the Eclipse platform and Sirius Editor, MAGDA exemplifies the advantages of using extensible frameworks that allow for the customization of modeling tools tailored to specific research needs. This seamless integration minimizes disruptions to modelers' workflows, thereby reducing the learning curve.

User experience is a critical factor in the design of such tools. To meet the requirements of a modeling tool enhanced with intelligent assistance, we implemented advanced features, including a multi-threaded architecture, caching strategies, and back-off mechanisms for efficient API call management. The careful selection of these technologies was instrumental in ensuring the tool's effectiveness. Additionally, the event-driven architecture ensures that each user action automatically triggers the necessary logging and, when appropriate, activates the prediction model to produce timely recommendations. This responsiveness is crucial for maintaining an engaging and productive modeling session, where the tool's assistance is intuitive and anticipates the modeler's needs.

Looking ahead, the integration of LLMs into modeling environments represents a natural next step for advancing intelligent assistance in modeling tools. These AI-models, capable of generating domain-specific recommendations, could significantly enhance the precision and creativity of design suggestions. We are currently developing a new version with locally executed LLMs to reduce response times. The increasing performance of these models, even with a relatively low number of parameters, makes this feasible.
 
More generally, we designed MAGDA to accommodate extended features and are continuously enhancing the user experience based on participant feedback. This iterative approach ensures that MAGDA remains an effective and user-friendly modeling tool, staying at the forefront of advancements in both Model-Driven Engineering and Artificial Intelligence.

\section{Conclusions}\label{sec:conclusion}

In this paper, we first presented a tool for domain modeling assistance that leverages LLMs to suggest domain elements. Through a user study, we then evaluated the utility of providing assistance during the domain modeling task using three different modes: automated suggestions, which provide continuous, real-time assistance; suggestions on request, which allows users to seek help as needed;  and suggestions at the end, which offer a list of missing elements after the initial model is completed.
Our findings indicate that this assistance significantly reduces the time required to complete modeling tasks, thereby boosting productivity. Importantly, offering suggestions throughout the modeling process not only increases their likelihood of acceptance but also enhances their impact on the final models. While the introduction of LLM-based assistance contributes positively to the modeling process, it also appears to reduce the diversity of models produced by different individuals within the same domain. This reduction could potentially restrict the creative input of modelers, although the current data does not conclusively confirm this effect. Participant feedback further highlights the value of the LLM-based tool, though some aspects of the user experience still need enhancement to better meet modeler needs.

This study shapes our vision of the future of modeling, where assistance powered by foundation models like LLMs plays a pivotal role in capturing and refining domain-specific problems during software development. As the capabilities of these foundation models evolve, we anticipate that intelligent assistance will not only expedite the modeling process but also improve the accuracy and consistency of models across diverse domains. However, for this vision to be fully realized, future tools must strike a balance between automating routine tasks and preserving the creative flexibility of modelers. By refining how suggestions are integrated into the modeling workflow, domain modeling should become more intuitive, efficient, and collaborative, driven by adaptive, context-aware assistance.

For future work, we aim to refine our LLM-based approach to generalize it to a variety of domain-specific modeling languages. Our goal is to systematically bridge the gap between the general knowledge encoded in LLMs and the specific rules and requirements of modeling formalisms. By moving beyond ad-hoc solutions, we will develop a more structured methodology that extends the applicability of our LLM-based assistance to encompass not only domain models but also a broad spectrum of static and dynamic software modeling formalisms.

\bibliographystyle{ACM-Reference-Format}
\bibliography{bib/references-clean}

\clearpage

\appendix

\onecolumn

\section{Questionnaire}\label{app:questionnaire}

\noindent 
\begin{minipage}{0.5\textwidth}
    \fbox{%
        \includegraphics[width=0.95\textwidth, page=1]{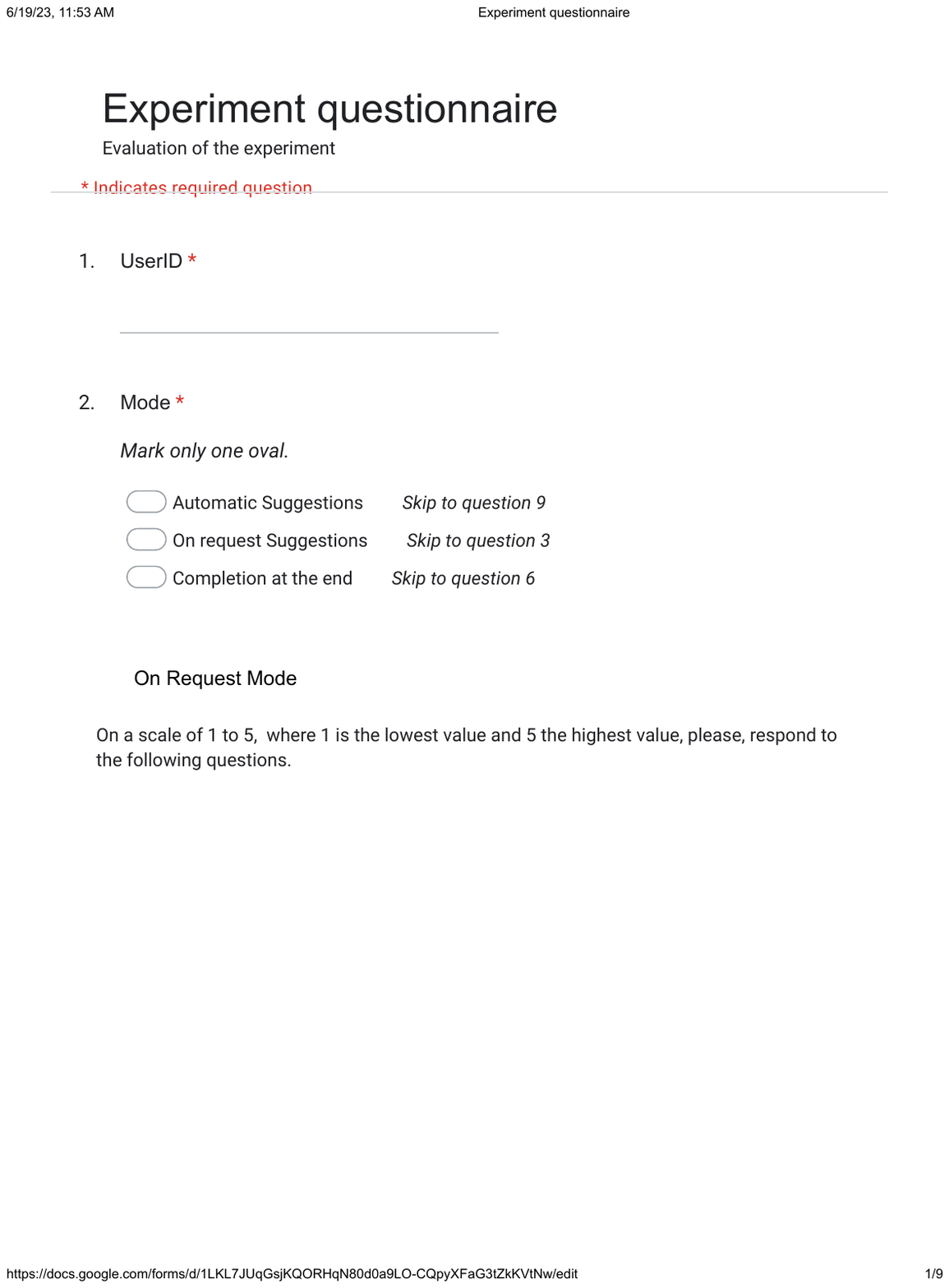}
    }
\end{minipage}%
\begin{minipage}{0.5\textwidth}
    \fbox{%
        \includegraphics[width=0.95\textwidth, page=2]{appendices/ExperimentquestionnaireGoogleForms.pdf}
    }
\end{minipage}

\vspace{5mm} 

\noindent
\begin{minipage}{0.5\textwidth}
    \fbox{%
        \includegraphics[width=0.95\textwidth, page=3]{appendices/ExperimentquestionnaireGoogleForms.pdf}
    }
\end{minipage}%
\begin{minipage}{0.5\textwidth}
    \fbox{%
        \includegraphics[width=0.95\textwidth, page=4]{appendices/ExperimentquestionnaireGoogleForms.pdf}
    }
\end{minipage}
\clearpage

\end{document}